\documentclass[galaxies,article,submit,pdftex,moreauthors,numbers]{Definitions/mdpi} 
\let\linenumbers\relax
\usepackage{comment}
\usepackage{epsfig}
\usepackage{epstopdf}
\usepackage{ulem}

\def\subfigure#1{#1}

\let\citealt\citet
\let\citenum\citep
\renewcommand{\citep}[1][]{\def\tmpa{#1}\citepcont}
\newcommand{\citepcont}[2][]{\def\tmpb{#1}(\ifx\tmpa\empty\else\tmpa\ \fi\citet{#2}\ifx\tmpb\empty\else\ \tmpb\fi)}

\def\apjl{ApJL}

\def\apjs{ApJS}

\def\aap{A\&A}

\def\sun{\odot}

\firstpage{1} 
\pubvolume{1}
\issuenum{1}
\articlenumber{0}
\pubyear{2022}
\copyrightyear{2022}
\datereceived{} 
\dateaccepted{} 
\datepublished{} 

\hreflink{https://doi.org/} % If needed use \linebreak
\Title{A Review of the Mixing Length Theory of Convection in 1D Stellar Modeling}

\TitleCitation{Title}

\Author{Meridith Joyce $^{1,2,\dagger}$\orcidA{}, and Jamie Tayar $^{3,\dagger}$\orcidA{}}

\AuthorNames{Meridith Joyce and Jamie Tayar}

\AuthorCitation{Joyce, Meridith; Tayar, Jamie}
\address{%
$^{1}$ \quad Konkoly Observatory, Research Centre for Astronomy and Earth Sciences, H-1121 Budapest Konkoly Th. M. \'ut 15-17., Hungary; meridith.joyce@csfk.org\\
$^{2}$ \quad CSFK, MTA Centre of Excellence, Budapest, Konkoly Thege Mikl\'os \'ut 15-17., H-1121, Hungary;\\
$^{3}$ \quad University of Florida; jtayar@ufl.edu\\
$\dagger$ \quad These authors contributed equally to this publication
}

\abstract{We review the application of the one-dimensional Mixing Length Theory (MLT) model of convection in stellar interiors and low-mass stellar evolution. 
We summarize the history of MLT, present a derivation of MLT in the context of the 1D stellar structure equations, and discuss the physical regimes in which MLT is relevant.
We review attempts to improve and extend the formalism, including to higher dimensions. We discuss the interactions of MLT with other modeling physics and demonstrate the impact of introducing variations in the convective mixing length, $\alpha_{\text{MLT}}$, on stellar tracks and isochrones. We summarize the process of performing a solar calibration of $\alpha_{\text{MLT}}$ and the state-of-the-art on calibrations to non-solar targets. We discuss the scientific implications of changing the mixing length, using recent analyses as demonstration. We review the most prominent successes of MLT and remaining challenges, and we conclude by speculating on the future of this treatment of convection.}


\begin{document}

%%%%%%%%%%%%%%%%%%%%%%%%%%%%%%%%%%%%%%%%%%

\section{Introduction}
\label{sec:introduction}
The matter of energy transport in stars is notoriously complicated. In particular, the details of convection in the stellar interior are difficult to probe with direct observation and encompass the dominant sources of uncertainty in stellar models \citep{Choi2018,
Tayar2022, Cinquegrana2022, Joyce2023}. Even the most specialized one-dimensional (1D) stellar evolution codes must simulate the behavior of stars over enormous ranges in temperature, density, and pressure as well as over evolutionary timescales.
{Simulations and measurements of convection in non-astrophysical cases indicate the importance of both large and small scale motions, and the need to adequately resolve both. When applied to astrophysical problems, which involve timescales ranging from the duration of nuclear processes (fractions of milliseconds) to tens of billions of years and spatial scales ranging from nuclear cross-sections to thousands of solar radii, this rapidly becomes intractable. Because of the ranges involved, accurately simulating all of these scales simultaneously is computationally infeasible.}
These requirements therefore demand a simplistic framework for convection. Though convection is an intrinsically three-dimensional, turbulent, non-linear, and time-dependent process, the practicalities of stellar modeling demand that we parameterize convection in a static and one-dimensional way. Mixing length theory (MLT) provides a solution.

The mixing length theory of convection describes the bulk movement of fluids in analogy with molecular heat transfer. 
Considering a pocket within a convection zone as a
discrete ``parcel'' of fluid with locally uniform physical characteristics, we may trace its vertical displacement. Assuming the parcel is in pressure equilibrium with its surroundings but not in thermal equilibrium, a relatively hot parcel will move towards a cooler region, and cooler parcels will move towards hotter regions. As hot parcels are under-dense relative to their surroundings, it is a buoyancy force that causes them to rise and expand. Cooler parcels, on the other hand, sink and compress.
The characteristic distance over which such a parcel can travel (along a radial line, in one dimension) before losing its locally homogeneous physical characteristics can be thought of as the \textit{mean-free path} of that parcel, measured in terms of the pressure scale height, ${\mathrm d} \ln(P)/{\mathrm d} \ln(T)$, of the stratified fluid. 

First applied to stellar interiors more than 60 years ago in an influential paper on solar convection \citep{BoehmVitense1958}, the mixing length theory of convection remains the dominant framework for one-dimensional (1D) convective energy transport used in stellar structure and evolution calculations. This longevity speaks not only to its robustness and effectiveness as a formalism, but also to the difficulty of constructing viable alternatives, even in today's era of exceedingly greater computing resources. With rare exception, all modern 1D stellar structure and evolution programs use MLT, or a close variant thereof, for convective energy transport.
Popular stellar evolution tools for the low- and intermediate-mass regimes include
%These include:
\begin{itemize}
\item[.] ATON Rome Stellar Evolution Code (\citealt{Ventura98, Ventura13});
\item[.] Bag of Stellar Tracks and Isochrones (BaSTI; \citealt{Pietrinferni04large}); 
\item[.] Cambridge STARS \citep{Eggleton71evolution};
\item[.] Code d'Evolution Stellaire Adaptatif et Modulaire (CESAM; \citealt{MorelLebreton2008});
\item[.] the Dartmouth Stellar Evolution Program (DSEP; \citealt{Dotter08});
\item[.] the Garching Stellar Evolution Code (GARSTEC; \citealt{Weiss08garstec});
    \item[.] the Geneva Stellar Evolution Code (GENEC; e.g.\ \citealt{Charbonnel96grids});
\item[.] Modules for Experiments in Stellar Astrophysics (MESA; \citealt{MESAVI, Paxton19instrument5, Paxton18instrument4, Paxton15instrument3, Paxton13instrument2, Paxton10instrument1});
\item[.] the Monash stellar evolution code (an adaptation of the Mount Stromlo Stellar Evolution code; \citealt{LattanzioThesis, Lattanzio86, Frost96, Karakas07}); 
\item[.] the PAdova and TRieste Stellar Evolution Code (PARSEC; \citealt{Bressan12}); and
\item[.] the Yale Rotating Stellar Evolution Code (YREC; \citealt{Demarque08yrec}),
\end{itemize}
Among these, only ATON and CESAM provide an alternative to MLT: the full spectrum of turbulence model of \citet{Canuto91} and \citet{Canuto96}, discussed further in Section \ref{sec:alternative_formalisms} (for a more thorough overview of stellar evolution codes and their specializations, see the introduction of \citealt{Cinquegrana2022}).

Driven in part by ambitious space-based surveys such as Gaia \citenum{GaiaEDR3}, TESS \citenum{TESS}, and Kepler \citenum{Borucki2010} and ground-based efforts such as LSST \citenum{LSST}, APOGEE \citenum{Majewski17}, LAMOST \citenum{Cui2012}, GALAH \citenum{GalahDR3}, and so forth, and the rich data climate they are generating, there is renewed interest in and understanding of the importance of model-derived fundamental stellar parameters. Given the ubiquitous use of MLT in stellar evolution calculations, we now present a review of 1D convection and its applications, especially to the theory and observation of low-mass ($\sim0.5$--$1.0M_{\odot}$) stars. 
%

%%%%%%%%%%%%%%%%%%%%%%%%%%%%%%%%%%%%%%%%%%
\section{History}
\label{sec:history}
Here we summarize the key milestones in the development and extension of the mixing length theory in its application to stellar interiors. %For more extensive reviews on 

In 1925, fluid dynamicist Ludwig Prandtl developed a simplified model of Reynolds stress in analogy with molecular heat transfer \citep{Prandtl25}. This set the precedent for describing turbulent motions using a diffusion approximation, which is a defining feature of all modern-day mixing length theory formulations. Almost 30 years later, Erika B\"ohm-Vitense developed the first incarnation of mixing length theory for use in models of stars and applied the framework to solar convection \citep{Vitense53}. Published 70 years ago at the time of writing, \textit{Die Wasserstoffkonvektionszone der Sonne}, or ``The Hydrogen Convection Zone of the Sun,'' remains the canonical reference for MLT treatments in models of the Sun and other stars, but it was not until five years later that she generalized her theory to stars with different effective temperatures in \citet{BoehmVitense1958}.

Her work was extended by Henyey, Vardya and Bodenheimer \citep{Henyey65mlt}, who focused on the formal theory for representing the superadiabatic layers in convective envelopes. This yielded a modified MLT formalism that still remains the optimal choice for use in optically thin regimes in stellar structure and evolution calculations. Following this, \citet{Cox68} presented an MLT formalism that was ideal for optically thick material. 

In 1971, Erika B\"ohm-Vitense's husband, Karl-Heinz B\"ohm, and J.\ Cassinelli studied MLT in the context of the thin convective envelopes found on white dwarfs \citep{BohmCassinelli1971}. Later that decade, Dmitri Mihalas \citep{Mihalas1978, Mihalas1978book} and \citet{Kurucz1979} published studies of radiative transfer models of stellar atmospheres in connection to sub-surface convection zones characterized by MLT. 

The advent of helioseismology (e.g. \citealt{Basu1994, Basu2000}) allowed for deeper understanding of the Sun's convection zone. In the late 90s, \citet{Demarque1997,Demarque1999} investigated the superadiabaticity of the Sun and the effects of MLT on the predicted acoustic pressure modes (or $p$-modes) propagating in the convective cavity. To this day, $p$-mode asteroseismology remains one of the most powerful tools for characterizing the outer envelopes of low-mass stars and a powerful diagnostic for our theoretical treatments of convection.

\section{Stellar Structure Context}
\label{sec:stellar_structure}
Here we briefly summarize the principles of stellar structure and evolution and the thermodynamic quantities necessary to build an intuitive picture of mixing length theory.

\subsection{Stellar Structure Equations}
In stellar evolution calculations, the equations of stellar structure are solved under the assumptions of conservation of momentum, conservation of mass (or \textit{mass continuity}), and hydrostatic equilibrium.
In the simplest scenario, it is assumed that hydrostatic equilibrium is achieved through the balance of gravitation against pressure. The Eulerian formulation\footnote{meaning the one-dimensional distance element is taken to be the fractional \textbf{radius}, $dr$, rather than mass, $dm$} of the canonical stellar structure equations is given by Equations \ref{11a}--\ref{11h}:
\begin{subequations}
\begin{alignat}{2}
\rho \frac{d {\bf v}}{d t} &= -\nabla P - \rho \nabla \Phi &&\qquad \text{ Momentum conservation } 
\label{11a}
\\
\nabla^2 \Phi &=  4 \pi G \rho \quad&&\qquad \text{ Gravitation }
\label{11b} 
\\[12pt]
\frac{d P}{d r}  &= - \rho \frac{d \Phi}{d r} &&\qquad \text{ Hydrostatic equilibrium} 
\label{11c}
\\
\frac{d M}{d r}  &= 4\pi r^2 \rho &&\qquad \text{ Mass continuity} 
\\
\frac{dL}{dr} &= 4\pi r^2 \rho \epsilon(\rho, T, \mu) && \qquad \text{ Conservation of energy }  
\\
\frac{d T}{d r} &= -\frac{3}{16 \pi ac} \frac{\kappa \rho L}{r^2 T^3} && \qquad \text{ Radiative energy transport }
\label{11f}
\\
%               &= -\left(1 - \frac{1}{\gamma} \right)\frac{T}{P} \frac{dP}{dr} && \qquad \text{ Adiabatic convective energy transport }
               &= -\frac{GM}{c_p r^2}  && \qquad \text{ Adiabatic convective energy transport } 
\label{11g}
\\[12pt]
\frac{d \rho_i}{d t} &= Q_i &&\qquad \text{ Nuclear energy generation.} 
\label{11h}
\end{alignat}
\end{subequations}

At its core, a stellar structure and evolution code is a \textit{polytrope} solver. A polytrope is a self-gravitating gaseous sphere, and this physical system provides a reasonable, first-order approximation of an uncomplicated star. 
A self-gravitating, spherically symmetric ball of fluid can be characterized using the dimensionless Lane--Emden equation:
\begin{equation}
    \frac{1}{\xi^2}\frac{d}{d\xi} \left(\xi^2 \frac{d\theta}{d\xi}\right) + \theta^2 = 0,
\label{eq:LaneEmden}
\end{equation}
where $\xi$ can be taken as a proxy for $r$ and $\theta$ as a proxy for density ($\rho$). We seek a solution $\theta(\xi) \sim \rho(r)$.  
The Lane--Emden equation must be solved under the assumption of an \textit{equation of state} relating pressure, density, and temperature, one example of which is the ideal gas law.
However, temperature can be neglected in the simplest model. The polytropic equation of state usually takes the form $P = K\rho^{1+ \frac{1}{n}}$, where $n$ is the polytropic index and $K$ is a constant of proportionality. Obtaining $\theta(\xi) \sim \rho(r)$ yields the \textit{stellar profile}, which describes the distribution of matter in the star.  

Solving this equation provides a solution for the stellar structure, but to compute the \textit{stellar evolution}, we must have a temporal component. 
%Implicitly, all functions of $r$ are also functions of time, $t$. 
In a real equation of state (e.g. the ideal gas law), there is temperature dependence, and the thermodynamic state of the model changes from time step to time step according to \textit{nuclear energy generation}. Monitoring how the stellar structure solution---in particular, the outer boundary values for quantities such as effective temperature $T_\text{eff}$, luminosity, and radius---changes as a function of time provides us with \textit{evolutionary tracks}. 
The key ingredients a user must specify when generating an evolutionary track are the mass, composition, and mixing length of the model. {While mass and composition are physical quantities in a way the mixing length is not, all three are equally important in determining the model star's evolutionary and structural behavior.}

\subsection{Thermodynamic Quantities and Convective Stability Criteria}
\label{sec:thermo_and_conv_stability}
During each time step, the model's thermodynamic structure must be calculated. This requires knowledge of whether any given radial shell is \textit{stable against convection}. To evaluate convective stability, we check the Schwarzschild \cite{Schwarzschild1958} and/or Ledoux criterion \cite{Ledoux1947}.\footnote{For discussion of the difference between these criteria and where they are applicable, consider reading e.g. \citet{Gabriel2014,SalarisCassisi2017,Anders22schwarzschild}} The Ledoux criterion for dynamical stability is given by
\begin{equation}
 \nabla_\text{rad} < \nabla_\text{ad} + \left[\phi/  \delta \right] \nabla \mu,
 \label{eq:Ledoux}
 \end{equation}
where $\phi, \delta$ are the partial derivatives of density with respect to temperature and composition, respectively,  $\nabla_\text{rad}$, $\nabla_\text{ad}$ are the radiative and adiabatic temperature gradients, respectively, and $\nabla_\mu$ is the composition gradient.
Under the simplification of homogeneous chemical composition ($\nabla \mu \rightarrow 0$), this reduces to the Schwarzschild stability criterion: $\nabla_\text{rad} < \nabla_\text{ad}$ . 

When the applicable condition is satisfied, the zone being evaluated  
%governed by the equation of state (EOS) $\rho(P, T, \mu)$ 
is dynamically stable. Dynamically stable regions do not produce convective motions, and so the energy flux is carried out exclusively by radiation (or conduction) in these regimes. 

If the convective stability criterion is not met, however, convection will activate and share in the transport of flux (i.e.\ luminosity or energy). In cases of efficient convection (such as deep core convection---see Section \ref{sec:where_does_MLT_matter}), the flux is carried entirely by convection. Stellar evolution calculations invoke MLT in cases where carriage of the flux is shared by radiation and convection.  

In this latter case, a useful toy model for mixing length theory is
\begin{equation}
    F_\text{conv} = \frac{1}{2}\rho v c_P T \frac{\lambda}{H_P}(\nabla_T - \nabla_\text{ad})
\label{eq:Fconv}
\end{equation}
with 
\begin{equation}
    \alpha_{\text{MLT}} \equiv \frac{\lambda}{H_P},
\label{eq:alpha_defn}
\end{equation}
where $\rho, v, c_P$ represent density, velocity, and specific heat, respectively, and the final term captures the balance of the global temperature gradient, $\nabla_T$, against the adiabatic temperature gradient, $\nabla_\text{ad}$ (a more formal derivation is given in Section \ref{sec:derivation}).
The definition in Equation \ref{eq:alpha_defn} is that of the \textit{mixing length parameter}, or $\alpha_{\text{MLT}}$. This is a dimensionless parameter characterizing the "distance," measured in terms of the pressure scale height $H_P = {\mathrm d} \ln(P)/{\mathrm d} \ln(T)$, that a parcel of convective material can travel. 

This $\alpha_\text{MLT}$ can be thought of in many ways, including as the convective mean-free path (as discussed in Section \ref{sec:introduction}), as a measure of the \textit{convective efficiency}, or as a pseudo-physical quantity that captures the change in entropy from the base to the top of the convection zone {(see end of Section \ref{sec:where_does_MLT_matter})}. The quantity $F_\text{conv}$ is determined according to two things: the difference in temperature gradients and the value assigned for $\alpha_\text{MLT}$. The suppression or enhancement of surface convective flux can be modulated by $\alpha_\text{MLT}$: {larger values for $\alpha_\text{MLT}$ mean more flux is carried by convection}. 
Changing this value has been shown to impact the predicted surface properties of low-mass stars in ways that should not be ignored (see Section \ref{sec:scientific_impacts}). 

%\section{When Do We Use MLT in Stellar Models?}
\section{Where Does the Choice of $\alpha_\text{MLT}$ Matter in Stellar Models?}
\label{sec:where_does_MLT_matter}
\begin{figure}[H]
\includegraphics[width=\textwidth]{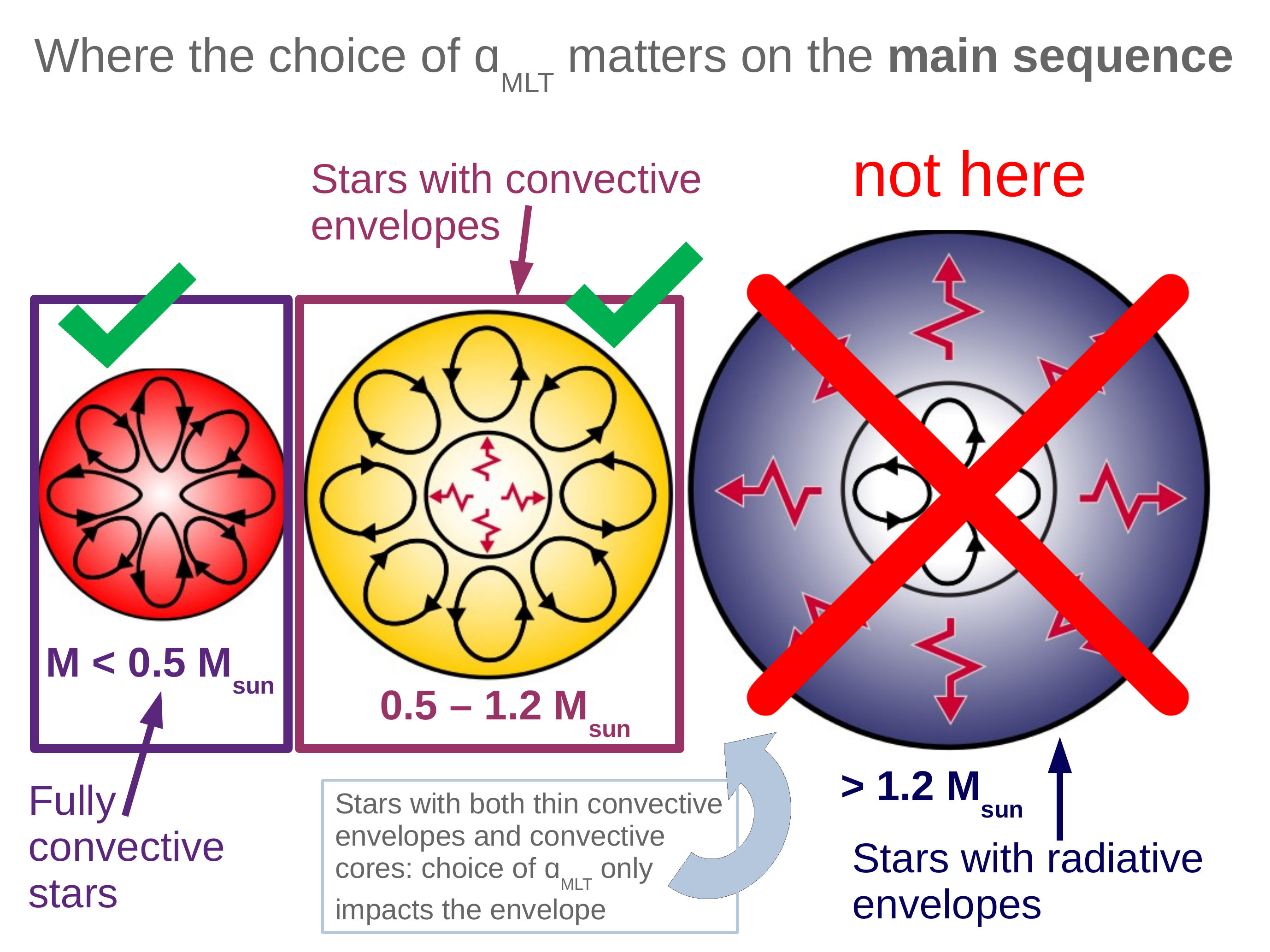}
\caption{Classification of stars according to convective structure on the main sequence. While all regions of the stellar model are treated using the MLT formalism, it is only fully convective stars and stars with convective envelopes (including those with both convective envelopes and convective cores) for which the \textit{choice} of $\alpha_\text{MLT}$ is relevant. This is because the temperature gradient is approximately adiabatic in, e.g., convective cores (see Section \ref{sec:derivation} for more detail).}
\label{fig:where_MLT_ms}
\end{figure}

Stellar modelers typically classify main-sequence stars according to mass, with three main categories: those that are fully convective (stars less than about $0.5 M_{\odot}$; e.g.\ M dwarfs), those that have radiative cores and convective envelopes ($0.5$--$1.2 M_{\odot}$, or stars like the Sun), and those that have convective cores with radiative envelopes ($> 1.2 M_{\odot} $). 
Figure \ref{fig:where_MLT_ms} is a cartoon depicting the ``convective configurations'' of main sequence stars of various mass. 
Mixing length theory is used to characterize energy transport in stars of all masses and all regions therein, but the \textit{choice} of $\alpha_\text{MLT}$ is only important for models in the lower two mass categories, and not the third.
It is also important in models of, e.g., F-type stars, which host a thin convective envelope as well as a convective core, with a radiative zone between them. In this case, the value of $\alpha_\text{MLT}$ is only relevant to the outer convection zone. 

The impact of $\alpha_\text{MLT}$ in mixing length theory is felt only in \textit{superadiabatic} regions, or regions where the temperature and density gradients align such that higher temperatures correspond to lower densities, causing hot material to rise. This is because the mixing length parameter is tied to a temperature excess that only exists in regions where the flux is carried by a combination of radiation and convection (whereas the flux is carried entirely by convection in the case of convective stellar cores, where the temperature gradient is nearly adiabatic; see Section \ref{sec:derivation} for more detail). 
%\blu{1 sentence on what happens in core convection}
%

Superadiabatic regions of low-mass stars include surface convection zones, sub-surface convection zones, and (later in their evolution) red giant envelopes. In intermediate- to high-mass stars, conditions may also produce intershell convection, or localized, interior pockets of convection situated between the core and the envelope.

Importantly, convection in the stellar core is not a case of superadiabatic convection. While convection at the surface is very inefficient, conditions in the core locally approximate an \textit{isentropic}\footnote{That is, where entropy is constant.} environment, meaning core convection is \textit{adiabatic}. 
The standard prescription for energy transport by convection in adiabatic regimes is given by Equation \ref{11g}. This can be written (in Lagrangian form, i.e.\ $m$ rather than $r$ coordinates) as
\begin{equation}
      \frac{d T}{d m}  = -\frac{T}{P} \frac{Gm}{4\pi r^4} \nabla  
      %&& \qquad \text{ Adiabatic convective energy transport }
\end{equation}
{where $\nabla$ represents the \textit{general} temperature gradient} (see Equation 7.32 and surrounding discussion in \citealt{Kippy}).
Note that the form of $\nabla$ is key: in the deep interior, $\nabla  = \nabla_\text{ad}$. In the regions where MLT applies, $\nabla$ is instead given by the solution to Equation \ref{eq:Fconv}. 

Referring again to Equation \ref{eq:Fconv}, we see how mixing length theory's key parameter, $\alpha_{\text{MLT}}$, can be conceptualized as a measure of convective efficiency, with higher values corresponding to the statement that a larger amount of flux is carried by convection. Changes to the stellar structure induced by changes in $\alpha_\text{MLT}$ will therefore be most significant when convective efficiency is low. 

Within a superadiabatic convective region, there is still a differential in the efficiency of convection as a function of depth\footnote{More precisely, it is the lack of an efficiency differential in the convective core that makes the choice of $\alpha_\text{MLT}$ irrelevant
in this regime}. While the deepest portion of the outer convection zone is nearly adiabatic, or \textit{asymptotically adiabatic}, the top of the convective envelope is not. The difference in entropy between these regions is also captured by $\alpha_\text{MLT}$.

\section{Limitations and Physics Not Captured by MLT}
\label{sec:limitations}

Mixing length theory requires a number of na{\"ive}---and in some cases, outright incorrect---physical assumptions. First of all, it models an \textit{advective} process using a diffusive approximation. Advective processes are those that transport material or energy through the bulk motion of fluids, whereas MLT supposes that particle-like fluid parcels diffuse through the region to redistribute heat. In the physical world, fluids and particles behave quite differently, and radiative hydrodynamics has shown that the concept of a homogeneous unit of convective material sustained over any appreciable distance is not valid. Convection proceeds via a continuum of constantly changing upflow and downflow channels.
Likewise, MLT treats convective boundaries as if they are rigid. In reality, convective boundaries are permeable, flexible, and subject to the inertia of convective motions carrying convective plumes across the convective--radiative boundary. Local mixing occurs in these regions, and MLT cannot account for this.

Another simplification supposed by MLT is that fluid parcels travel along strictly vertical paths---this limitation is primarily due to the 1-D aspect of the formulation. In physical convection, there is continuous shearing, fragmentation, reorientation, and deletion of the flow channels. None of these features can be captured using a formulation that relies on strictly radial displacements. It should also be emphasized that standard MLT is 
\textbf{not} time-dependent, which means it cannot capture any physics happening faster than the convective turnover time. In standard treatments, convective regions are assumed to be instantaneously mixed over one evolutionary time step.

On the issue of flow channels, MLT fails to account for 
%\textit{negative kinetic flux}, which is an important part of physical convection emerging from 
the fact that, in physical convection, there is asymmetry between upflows (hot material rising) and downflows (cool material falling). Taking, for example, the surface of the Sun or 3D simulations thereof, we observe that a network of broad, spot-like convective cells with higher temperatures is demarcated by a series of interconnected, cooler downflow lanes. The surface area is dominated by plasma flowing upward, which expands as its density drops, and this material travels in the same direction as the density gradient. However, the same is not true of the down-flowing material: this travels against the density gradient and generates turbulence in the process. To satisfy both conservation laws and the density gradient, the upflows lose mass to the downflows, and the downflows accumulate contributions from the upflow lanes at many different radial and density coordinates. 

Because the upflows carry material that is uniformly from the deep interior of the convection zone to the surface, the process is largely isentropic. 
There is no such uniformity in the downflow lanes, however: the material in these lanes has a range of entropies, and it is also denser. The downflows therefore cause turbulence, have higher speeds than the upflows, and occupy a smaller area than the upflows. These conditions result in the inward (towards the interior) transfer of kinetic energy, a process known as \textit{negative kinetic flux}. Negative kinetic flux is a physical property of convective plasma that classical MLT cannot capture, though some modern extensions and revisions of MLT have attempted to incorporate this feature (see \citealt{SteinNord89} for a detailed discussion of 3D plasma physics).

It is likewise important to recognize that the mixing length parameter, $\alpha_{\text{MLT}}$, of most MLT formulations is only loosely connected to any physical property of a star; it is often thought of as a free, numerical parameter. The mixing length is, however, related to the entropy jump between the asymptotically adiabatic portion of the convection zone (where convection is most efficient) and the top of the convective envelope (where it is least efficient). 
Although neither this entropy jump nor the density gradient, size, or depth of the convective envelope is a readily observable feature of a star, these features can be probed indirectly using asteroseismology. It is thus necessary to calibrate $\alpha_{\text{MLT}}$ directly, and this is most easily done using observations of our nearest star---a point to which we return in Section \ref{sec:solar_calibration}.

\section{Mixing Length Formulation} 
\label{sec:derivation}
%https://www.authorea.com/doi/full/10.22541/au.163837787.77365607/v1
%https://ui.adsabs.harvard.edu/abs/2022ApJS..262...19J/abstract

We reproduce here a standard derivation for the amount of flux carried by convection, modified from derivations presented in Cox \& Giuli's \textit{Principles of Stellar Structure}, Kippenhahn and Wieger's \textit{Stellar Structure and Evolution}, Cassisi \& Salaris' \textit{Stars and Stellar Populations}, and notes compiled by Matteo Cantiello and Yan-Fei Jiang (\textit{priv. comm.}).

The pressure scale height, $H_p$, is a measure of the distance over which the total pressure, $P = P_\text{gas} + P_\text{rad}$, changes by a factor of $1/e$. Under the assumption of hydrostatic equilibrium, the definition 
\begin{equation}
    -\frac{ \text{d} \ln P}{ \text{d} r} \equiv \frac{\rho g}{P} = \frac{1}{H_P}
\label{eq:Hp}
\end{equation}
holds.

\begin{figure}[H]
\includegraphics[width=\textwidth]{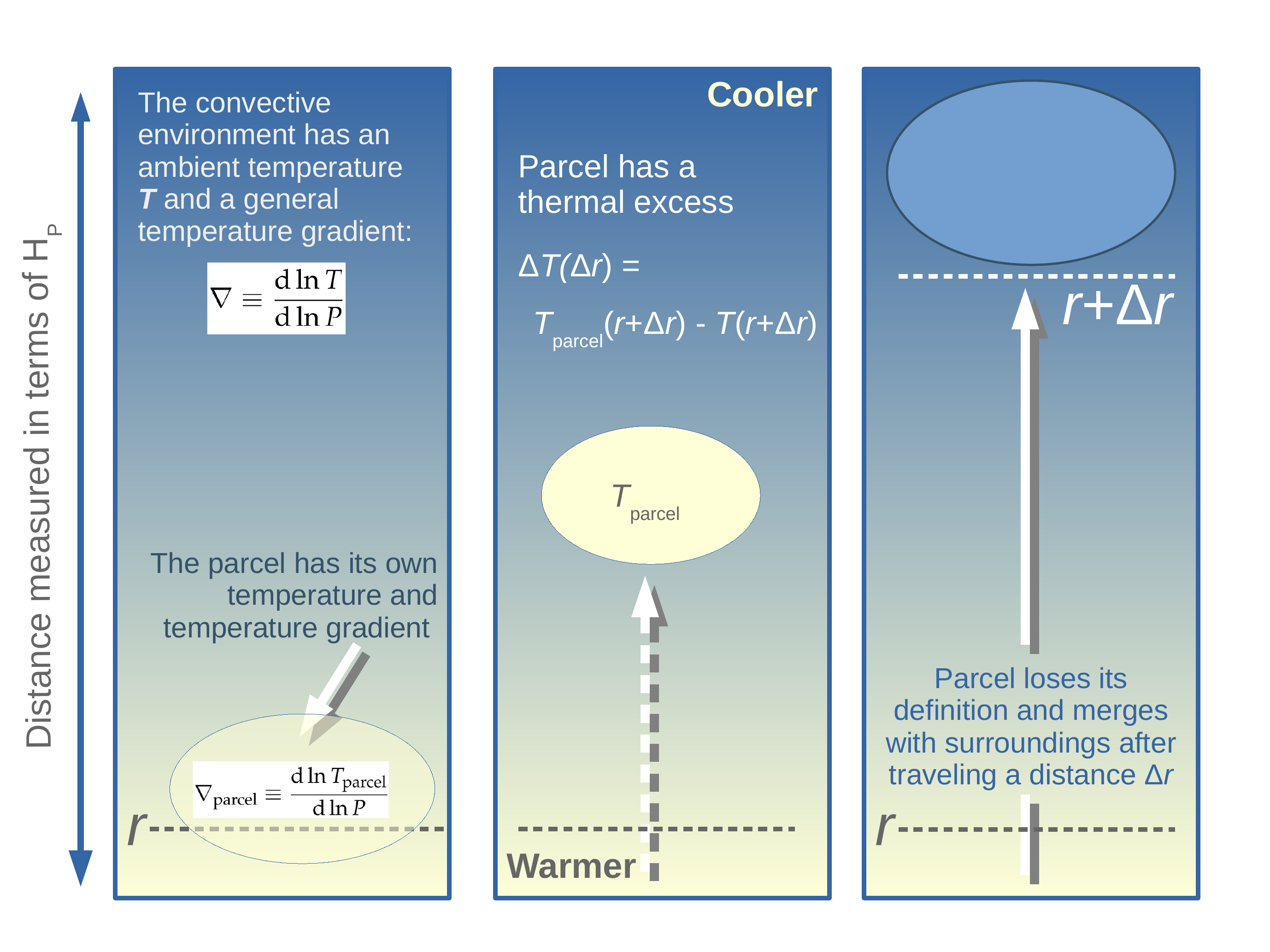}
\caption{The distance a "convective blob" can travel is measured in multiples of the pressure scale height, $H_P$. The upward motion of the parcel is driven by the thermal excess the fluid parcel has compared to its surroundings. A larger mixing length implies that the parcel travels over a larger pressure differential before denaturing,  which corresponds to more efficient transport of the flux by convection. 
%The Mona Lisa of our time. Options 1 and 2}
}
\label{fig:fluid_parcel}
\end{figure}

MLT's canonical ``parcel of fluid'' is taken to be in pressure, but not thermal, equilibrium with its surroundings.
We next define (1) the average (ambient) temperature gradient of the fluid with respect to the pressure of all matter at some $r$, and (2) a temperature gradient of the fluid parcel itself, also taken with respect to the total pressure. 
Let the former be given by 
\begin{equation}
    \nabla \equiv \frac{ \text{d}\ln T}{ \text{d} \ln P}
\label{eq:grad_total}
\end{equation}
and the latter by
\begin{equation}
    \nabla_\text{parcel} \equiv \frac{ \text{d}\ln T_\text{parcel}}{ \text{d} \ln P},
\label{eq:grad_parcel}
\end{equation}
where $T$ and $ T_\text{parcel}$ are the average (ambient) temperature and the temperature of the parcel, respectively.
{This scenario is depicted in Figure \ref{fig:fluid_parcel}.}

By first-order Taylor expansion, the difference between the parcel's temperature and the ambient temperature at some radial shift $\Delta r$ can be expressed as
\begin{equation}
    \Delta T(\Delta r) = 
    T_\text{parcel}(r + \Delta r) - T(r + \Delta r) \simeq \Delta r \left[ \frac{ \text{d} T_\text{parcel}}{ \text{d} r} - \frac{ \text{d} T}{ \text{d} r}  \right].
\end{equation}
Assuming that the temperature change over $\Delta r$ is small,  $T \simeq T_\text{parcel}$, and so
\begin{equation}
    \Delta r \left[ \frac{ \text{d} T_\text{parcel}}{ \text{d} r} - \frac{\text{d} T}{ \text{d} r}  \right] \rightarrow 
    \Delta r T \left[ -\frac{ \text{d} \ln T}{ \text{d} r} - \left(- \frac{ \text{d} \ln T_\text{parcel}}{ \text{d} r} \right) \right].
\label{eq:dlnT}
\end{equation}
Using the chain rule and the definition from Equation \ref{eq:Hp}, we can rewrite
\begin{equation}
\frac{\text{d} \ln T}{ \text{d} \ln P} = 
\frac{\text{d} \ln T}{ \text{d}  r} \frac{\text{d} r}{ \text{d} \ln P} = \frac{\text{d} \ln T}{ \text{d} r} \left(- H_P \right).
\label{eq:chain_rule}
\end{equation}
We may now substitute the definitions from Equations \ref{eq:grad_total} and \ref{eq:grad_parcel} into Equation \ref{eq:dlnT}, yielding
\begin{equation}
    \Delta T (\Delta r) = \Delta r \frac{T}{H_P}(\nabla - \nabla_\text{parcel}). 
\label{eq:deltaTdeltar}
\end{equation}
We note that the moving parcel may well exchange heat with its environment, and so $\nabla_\text{parcel}$ is a function of the rate at which this exchange takes place. 
However, the assumption that the parcel does \textit{not} exchange heat with its surroundings---i.e.\ that it is moving \textit{adiabatically}---is a reasonable and simplifying assumption for stellar conditions. In this case,
\begin{equation}
    \nabla_\text{parcel} \rightarrow \nabla_\text{ad} = 
   % \sout{ \left( \frac{ \text{d} \ln T}{ \text{d} \ln P}  \right)|_\text{ad}}
    \left. \frac{ \text{d} \ln T}{ \text{d} \ln P}  \right|_\text{ad}
    .
\label{eq:grad_ad}
\end{equation}
Take careful note of the difference between Equations \ref{eq:grad_ad} and \ref{eq:grad_total}: these are \textbf{not} interchangeable. The adiabatic temperature gradient, $\nabla_\text{ad}$, and the average or ambient temperature gradient $\nabla$ are not, in general, the same.  We also do not necessarily know the definition of $\nabla$ in a convective region. 
Substituting $\nabla_\text{parcel}$ for $\nabla_\text{ad}$ and combining with Equation \ref{eq:deltaTdeltar} yields

\begin{equation}
    \Delta T (\Delta r) = \Delta r \frac{T}{H_P}(\nabla - \nabla_\text{ad}). 
\label{eq:almost}
\end{equation}

Meanwhile, the convective flux transported by a parcel of fluid moving upwards with velocity $v$ over some distance $\lambda$ is given by
\begin{equation}
    F_\text{conv} = \frac{1}{2} \rho v c_P \left[  \left(\frac{\text{d}T}{\text{d}r} \right)_\text{ad} - \left( \frac{\text{d}T}{\text{d}r} \right) \right] \lambda,
\end{equation}
where $\rho$ is  density, $c_P$ is specific heat, and the factor of $\frac{1}{2}$ emerges from the assumption that half of the material in a given layer is rising while half is falling
(see \citealt{SalarisCassisibook2006} for more discussion of how this formula arises). Given the relationship 
\begin{equation}
    \frac{\text{d}T}{\text{d}r} = -\frac{T}{H_P} \frac{\text{d}\ln T}{ \text{d}\ln P} =  -\frac{T}{H_P}\nabla
\end{equation}
and Equation \ref{eq:deltaTdeltar}, we have 
\begin{equation}
     F_\text{conv} = \frac{1}{2} \rho v c_P \left[  \left( - \frac{T}{H_P} \nabla_\text{ad} \right) - \left( -\frac{T}{H_P} \nabla \right) \right] \lambda.
\end{equation}
Rearranging, we see that Equation \ref{eq:Fconv}, 
\begin{equation*}
    F_\text{conv} = \frac{1}{2}\rho v c_P T \frac{\lambda}{H_P}(\nabla_T - \nabla_\text{ad}),
\end{equation*}
describing the toy model introduced in Section \ref{sec:stellar_structure}, emerges.

We return briefly to stars with convective cores, as discussed in Section \ref{sec:where_does_MLT_matter}: in convective cores, where convection is efficient, conditions are nearly isentropic and the general temperature gradient is very nearly an adiabatic temperature gradient. In this case, $(\nabla_\text{T} - \nabla_\text{ad})  \rightarrow (\nabla_\text{ad} - \nabla_\text{ad}) = 0$, so we can see that the choice of $\alpha_\text{MLT}$ does not matter in this situation.
%the mixing length formulation is not applicable. 

\subsection{Specific Formulations}
There are several implementations of MLT, some of which are more appropriate for particular physical scenarios. One of the most commonly used today is the \citet{Cox68} prescription, which assumes high optical depth and no radiative losses, as does the original formulation of \citet{Bohm-Vitense1958}. The methods of \citet{Henyey1964}, \citet{Mihalas1978} and \citet{Kurucz1997} all provide extensions of MLT to work in optically thin regimes. To use MLT in regimes where electron degeneracy and pressure ionization are relevant, as in the convective envelopes of white dwarfs, the \citet{BohmCassinelli1971} extension is most appropriate.

The use of time-dependent convection has its origins in the study of stellar pulsations, beginning with the work of \citet{Unno1967} and \citet{Gough1965} 
in adapting the mixing length formulation to model the interaction of the turbulent velocity field with radial pulsations (see \citealt{HoudekDupret2015} for a detailed review of this topic).
Recent innovations in time-dependent convection include implementing the model of \citet{Kuhfuss1986} in MESA \citep{MESAVI}, which reduces to the standard \citet{Cox68} treatment over long timescales.

%%%%%%%%%%%%%%%%%%%%%%%%%%%%%%%%%%%%%%%%%%%%
\section{Alternatives and Extensions}
The simplistic assumptions on which mixing length theory is based have long cried out for a more sophisticated and realistic formulation. Numerous authors have suggested potential changes to the mixing length formalism, including the inclusion of non-local elements, use of a more consistent picture of the turbulent dissipation, use of more physically motivated parameterizations, and/or  use of more complex calibrations. 

\subsection{Alternative 1D Formulations}
\label{sec:alternative_formalisms}

Improvements to MLT must balance increased sophistication against the need to remain implementable in one dimensional stellar evolution models, and many attempts have been made. However, while such modifications often provide better answers in particularly challenging regions of stellar evolution (e.g. time-dependent convection in late stages of nuclear burning for massive stars), 
they are often discussed only in the local context of a particular problem. The comparative simplicity and wide applicability of the standard mixing length formalism has thus far prevented any wide adoption of a significantly different alternative.
We discuss a few alternative formulations of the mixing length theory here, although we acknowledge that there is a wider body of work in this field than we can adequately document here \citep[see eg.][]{HoudekDupret2015}. 

One early challenge to the mixing length formalism was the fact that its purely local formulation meant that 
it did not accommodate convective overshoot: MLT provided no way to get convective regions to overshoot into adjacent, formally stable radiative regions, even though observations of real stars seemed to demand some such process \citep{Gough1977, Renzini1987, GrossmanNarayan1993}. 
Attempts have been made to recompute the theory in a non-local way by including additional convective terms \citep{Eggleton1983, Kuhfuss1986, Xiong1986, Xiong1989a, Xiong1989b, Grossman1996}, but such expansions rapidly develop a significant number of extra terms as they try to include higher-order effects. 

There are classes of modified mixing length models that can, in theory, address other challenges, including  time-dependent effects in a pulsating star \citep{Gough1977}, the impact of composition gradients, representation of the depth dependence of rotation and magnetism \citep{IrelandBrowning2018} and so forth. Practically, however, such work has been limited to specific problems addressed on an individual basis, rather than leading to a true overhaul of the underlying framework.

There have also been more explicit attempts to include a physical description of turbulence in the parameterization of convection. In the inviscid interior of a star, it is expected that a wide spectrum of turbulent eddies contribute to the convective flux \citep{Marcus1983,Canuto91}, rather than the single eddy assumed by the standard mixing length theory. This change in the treatment of energy transport corresponds to differences in the flux carried and the resulting temperature structure \citep{Canuto96}. This model, known as \textit{full-spectrum turbulence}, has been adopted in some stellar evolution codes, including CESAM \citenum{MorelLebreton2008} and ATON \citenum{ventura1998}. 

\subsection{Extensions to 3D}
More recently, as three-dimensional simulations of convection have improved, researchers have attempted
to use these simulations to constrain the mixing length in a way that more realistically represents the physics of convection. Such simulations can incorporate not only the local and static mixing length, but information on the global derivatives, asymmetries between upflows and downflows, the larger scale properties of the star, and transverse versus radial differences \citep{Arnett2015}. For reasons likely related to the physics of turbulent dissipation \citep{Arnett2018}, most such simulations tend to find that the value of the mixing length should vary only slightly with the composition, luminosity, and surface gravity of the star \citep{ Trampedach2014, MagicMLT,Sonoi2018}. However, some authors have attempted to better reproduce the temperature stratification (T-$\tau$) relations in the three dimensional simulations \citep{Tanner2014, SalarisCassisi2015, Mosumgaard2018, Zhou2020}. Still others have attempted to incorporate additional physical information
%\sout{more information from real physics}
by calibrating the mixing length parameter directly to the entropy profile \citep{SpadaDemarque2019, Spada2021}. 
% https://ui.adsabs.harvard.edu/abs/2021MNRAS.504.3128S/abstract

While trends between $\alpha_\text{MLT}$ and other global properties (e.g. metallicity) from 3D simulations often go in the same direction as the trends implied by calibrations of the mixing length in 1D stellar models to match observations of stars \citep{Bonaca2012, Creevey2015, Tayar2017, Joyce2018aNotAll, Joyce2018balphaCen, viani2018}, they rarely agree quantitatively. In particular, there is a large discrepancy in magnitude: 1D-to-observational calibrations suggest, in some cases, the need for a variation in $\alpha_\text{MLT}$ that is a factor of 10 larger than suggested by 3D simulations. One may compare, e.g., \citealt{Joyce2018balphaCen} to \citealt{Trampedach2014} on this issue.
This suggests either a need for improvement in three dimensional simulations---for example, by extending their temperature and density domains or by extending their timescales---or the need for additional corrections to the physics of 1-D stellar evolution calculations that are currently impacting the inferred mixing length \citep{Choi2018, Valle2019}. We expect both the 3D and 1D communities will continue their efforts to search for a consensus method that brings the two sets of models into agreement.

\section{Standard 1D MLT and Its Interplay With Other Modeling Physics} 

In addition to the challenges related to the mixing length by itself, stellar modeling relies upon a large number of other physical inputs, including formulations of convective overshoot, opacities, equations of state, the treatment of diffusion, nuclear reaction rates, and so on, {and the interplay between $\alpha_\text{MLT}$ and these other physical assumptions must be considered}. While entire articles can and have been be written on each of these choices, its associated uncertainties, and its impacts on stars of various masses and evolutionary states, our focus here is on how these other physical choices interact with  the mixing length.

\subsection{Atmospheric Boundary Conditions}
The creation of a stellar structure model is in essence a boundary value problem, where choices about the outer boundary condition will have a significant impact on the resulting solution. Unfortunately, the outer boundary is where radiation begins to escape the star. This is precisely where the assumptions of stellar interiors begin to break down, and analytic solutions accumulate error. 
Most modelers deal with this problem by assuming a  relationship between the temperature and optical depth in this region (i.e. a T-$\tau$ relation). This can be an analytical expression (e.g. Eddington, \citealt{KrishnaSwamy}, \citealt{Ball2021}) 
or a table of values sourced from more sophisticated stellar atmosphere calculations that better include some of the physics of radiative transfer and loss \citep{Kurucz1997, CastelliKurucz2004, Hauschildt1999a}. These choices change the structure of the model in the {same region that is affected by the mixing length} {superadiabatic outer layers of the star}, and so the chosen atmospheric boundary conditions will change the  mixing length. This effect is most obvious in red giant stars, due to their larger convective envelopes. The mixing length is sensitive to both the atmospheric relation adopted and the optical depth in the atmosphere at which the boundary is set \citep{Choi2018}. 

We note that stellar atmosphere calculations also require a choice of mixing length \citep{Gustafsson2008}, though the choice of the mixing length in the atmosphere models is generally found to have only marginal impact on their physical predictions (e.g., abundances inferred; \citet{Song2020}). However, the mixing length chosen in stellar atmosphere models is rarely forced to be the same as the mixing length chosen in the stellar evolution models. Generally, this inconsistency is assumed not to matter, and the choice to ignore it can be reconciled in a framework where the mixing length is viewed as a tuning parameter accounting for other physical inconsistencies. However, it is more difficult to ignore this inconsistency in a framework where the mixing length is thought to capture
something physical about convection.

There are also arguments that three dimensional simulations give a better estimation of the relationship between temperature and optical depth, and that these should be used as boundary conditions in lieu of the standard one dimensional atmosphere models \citep{Trampedach2014, Tanner2014, MagicMLT}. Attempts have been made to implement these boundary conditions in a way that takes into account the temperature, metallicity, and luminosity of the star \citep[e.g.][]{Mosumgaard2018, Mosumgaard2020}, but in general, the effects on the stellar temperature and assumed mixing length have been smaller than expected (generally on the order of tens of Kelvin; \citet{Tanner2014, Mosumgaard2020}).

\subsection{Convective Boundaries}
Designing stellar models requires assigning the conditions {for convection.
This involves a choice between the Schwarzschild and Ledoux criteria for convective stability (see Section \ref{sec:thermo_and_conv_stability}), as well as a choice in the mathematical and numerical approach to}
locating the convective--radiative boundary (or boundaries). While these choices are arguably most important for convective stellar cores (e.g. \citet{Pedersen2021}) \footnote{The treatment of core convective boundaries is particularly important for determining whether massive stars will meet the $M_\text{core}$ criterion for death as a supernova.},
they can nonetheless carry significant implications for convective envelopes as well---for example, dredge-up events and nucleosynthetic yields from TP-AGB stars are highly sensitive to these conditions \citep{Cinquegrana2022, Karakas2022}.
As such, there are important choices to be made about the physics of the lower boundary of the surface convective zone, where the local temperature gradient is not as close to adiabatic as the temperature gradient near the core and interactions with the mixing length can occur. 
This is the location of a process sometimes referred to as \textit{convective undershoot}.

 Early work with the mixing length formalism suggested that the Schwarzschild criterion was more likely to be appropriate for setting this boundary \citep[e.g.][]{GrossmanNarayan1993}.
 Further work has been consistent with that result, indicating that while boundaries tend to be instantaneously consistent with the predictions of the Ledoux criterion, over time they will mix in additional material \citep{Paxton19instrument5} through processes like entrainment or oscillatory double diffusive convection \citep{Mirouh2012} until they grow to the size predicted by the Schwarzschild criterion \citep{Anders22schwarzschild}.

\subsection{Opacities}
There are numerous physical assumptions that impact the detailed structure and temperature profiles of the model, meaning they change our expectations for a solar model and therefore the mixing length required to reproduce the properties of the Sun at the solar age (see Section \ref{sec:solar_calibration}). We discuss this here in the context of opacities, but an analogous discussion could be imagined for composition, including individual abundance variations \citep{Pietrinferni2009,Beom2016}, the treatment of diffusion \citep{vanSaders2012a}, and any similar assumption that affects the superadiabatic layers. 

Opacity in the stellar interior is generally treated in a simplistic way in which one assumes an abundance for each element present in the star and reads in a table corresponding to that composition. The table contains an estimate of the Rosseland mean opacity for the prescribed mixture as a function of density and temperature. This is then used to compute how much energy can be carried by radiation. {The radiative flux, in turn, helps determine} which regions will convect and what the temperature and density of those regions will be. 
This means that changing the opacities will result in small changes to the estimated radius and surface temperature of the star at any particular time. Since calibrations of the mixing length in stellar models are generally done to fit a particular radius and temperature at a particular time, calibrated models with different opacities will require different mixing lengths, and these differences in mixing length will propagate into differences in the evolutionary timescale, pulsations, and nucleosynthesis \citep{Cinquegrana22solarcal}. 

\subsection{Magnetic Fields}
Resolved studies of the solar surface make it clear that magnetism, and particularly the concentration of magnetic field, can impact the local properties of convection \citep[e.g.][]{Roudier2023}. This sort of spatially resolved behavior is, however, challenging to reproduce with one dimensional stellar evolution models. Originally, such effects were expected to be unimportant and therefore ignored, but that is no longer the case for all stars. In particular, large regions of concentrated magnetic field have been observed on a variety of young stars, on M dwarfs, and on rapidly rotating evolved stars, including the sub-sub-giants \citep[e.g.][]{Gosnell2022,Libby-Roberts2023}. It is now understood that magnetism in these regions influences the surface convective flux, and these effects can be parameterized in a way that mimcs the behavior of the mixing length (e.g. \citet{IrelandBrowning2018}).

The magnetic field is thought to act in some cases as an additional magnetic pressure or energy density term that blocks the transport of flux through certain spatial regions, and this effect has been parameterized in some 1D stellar evolution codes 
(e.g. YREC, \citet{Somers2015}; MESA, \citet{MESAVI}.)
One of the simplest ways to incorporate this effect is by lowering the effective mixing length of the star, which will tend to inflate the star in much the same way as surface magnetism. This inflation increases consistency with the larger-than-expected observed radii of M dwarfs in binary systems \citep{Chabrier2007, SomersPinsonneault2014}. 

More sophisticated analyses can self-consistently incorporate the effects of magnetism into a modified mixing length theory \citep{FeidenChaboyer2012b}, or add corrections for the spatial inhomogeneities of star spots \citep{Somers2020}, but all of these tend to alter the structure in a way similar to alterations to the mixing length directly. This highlights the complex nature of the mixing length; while clearly a nonphysical parameterization, its values may also point to physical changes in the properties of convection and its interactions with the physics of the stellar interior. 

\section{What Does Changing the Mixing Length Do in Stellar Models?}
\begin{figure}[H]
\includegraphics[width=0.8\columnwidth]{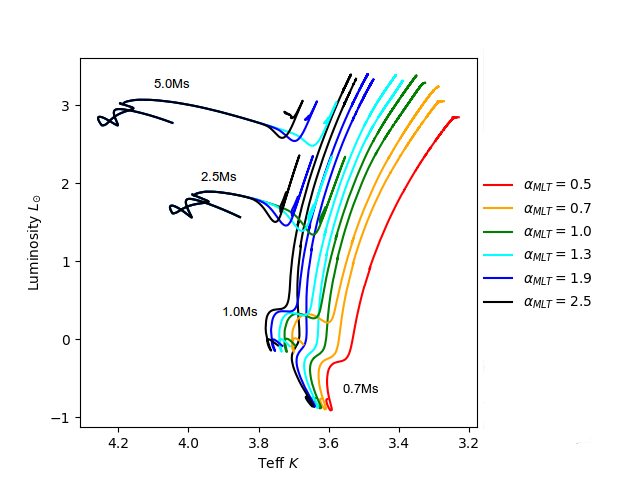}
\caption{Stellar tracks computed with the Dartmouth Stellar Evolution Program (DSEP) for a range of masses and mixing lengths. }
\label{fig:DSEP_mlt_variation}
\end{figure}

The effects of changing the value of $\alpha_{\text{MLT}}$ assigned in stellar evolution calculations performed with the Dartmouth Stellar Evolution Program (DSEP; \citealt{Dotter08}) are shown in Figure \ref{fig:DSEP_mlt_variation}. We note first of all that the effect of varying $\alpha_\text{MLT}$ is itself mass-dependent, with tracks at 2.5 or 5 $M_{\odot}$ showing no change along the main sequence with varying $\alpha_\text{MLT}$, whereas those at 1.0 and 0.7 $M_{\odot}$ show a shift towards cooler temperatures with decreasing $\alpha_{\text{MLT}}$. This is because, above a mass of approximately $1.2 M_{\odot}$, the structure of the star switches from hosting a convective envelope on the main sequence to hosting a radiative envelope, as explained in Figure \ref{fig:where_MLT_ms}. As changes to the mixing length will only affect superadiabatic regions with inefficient convection, changing $\alpha_{\text{MLT}}$ for models that only exhibit core convection will have no impact. However, we also observe that the effects of changing $\alpha_\text{MLT}$ begin to manifest for the higher-mass tracks after the main sequence turn-off. This corresponds evolutionarily to the development of a convective envelope, which introduces into the model a region of inefficient convection where MLT applies. We see here that lowering the value of $\alpha_\text{MLT}$ results in an extension of the subgiant phase, causing it to both lengthen in duration and shift towards cooler temperatures, for the 2.5 and 5.0 $M_{\odot}$ tracks. 

For the groups of 1.0 and 0.7 $M_{\odot}$ tracks, changes to $\alpha_{\text{MLT}}$ will have an impact throughout their main sequence, subgiant, and red giant evolutionary phases, as a convective envelope persists throughout these phases. Another observation is that changes to $\alpha_\text{MLT}$ are not linear: the temperature difference between tracks with $\alpha_\text{MLT}= 2.5$ vs $\alpha_\text{MLT}= 1.9$ is much smaller, in both mass cases, than the difference in temperature caused by a change from, e.g., $\alpha_\text{MLT}=0.5$ to $\alpha_\text{MLT}=1.0$. Intuitively, this is because the impact of changing the efficiency of convection is greatest when the efficiency is low.  

In Figure \ref{fig:MESA_mlt_variation}, we take a closer look at the $1M_{\odot}$ case. Tracks in this figure are computed with MESA rather than DSEP, and variations in the mixing length span $\alpha_\text{MLT} = 0.4$---a value suitable for some M dwarfs (cf.\ \citealt{Mann2015})---to $\alpha_\text{MLT}=2.1$, which is slightly greater than MESA's default (solar) value, $\alpha_\text{MLT, MESA default} = 2.0$. The same trends observed for the $1.0 M_{\odot}$ case in Figure \ref{fig:DSEP_mlt_variation} are visible here, including: 
\\(1) the shift towards cooler temperatures with lower $\alpha_\text{MLT}$; 
\\(2) the difference in effect of a 10\% change in $\alpha_\text{MLT}$ at, e.g., $\alpha_\text{MLT}= 2.1$ compared to $\alpha_\text{MLT}=0.8$;
\\(3) the extension of the subgiant branch in duration and towards cooler temperatures with decreasing $\alpha_\text{MLT}$; and
\\(4) the negligible impact on luminosity.
\\However, we note an additional feature in the tracks with the lowest mixing length values: a hook emerges near the end of the main sequence for $1 M_{\odot}$ tracks with $\alpha_\text{MLT} = 0.4$ and $0.5$. This corresponds to the development of a convective core---a feature that would not normally be present in a $1M_{\odot}$ model, but emerges in this case due to the suppression of convective flux at the surface of the model and the inward-propagation of those effects to the stellar interior. 
The impact of suppressing convective flux to this degree is significant enough to cause a structural realignment throughout the stellar model---specifically, an inflated radius---which 
results in a non-standard convective structure for the solar-like tracks with $\alpha_\text{MLT} < 0.6$.

\subsection{Impact on Isochrones}
Isochrones---from \textit{iso} meaning ``single'' and \textit{chronos} meaning ``time''---are models that represent a snapshot in time as a function of stellar mass. Whereas stellar tracks depict the time-evolution of a single star of particular mass, isochrones are constructed by interpolating over grids of stellar tracks, each of which has a different initial mass but otherwise identical input physics. Since stars of different masses evolve at different rates, points of equal age will occur during different evolutionary stages for different tracks. The curve connecting these equal-aged points across masses is the isochrone. For a more rigorous discussion of how these equal-aged points are defined, we refer the reader to \citealt{Dotter16eeps} for a description of the Equal Evolutionary Point, or EEP, method.

\begin{figure}[H]
\includegraphics[width=0.8\columnwidth]{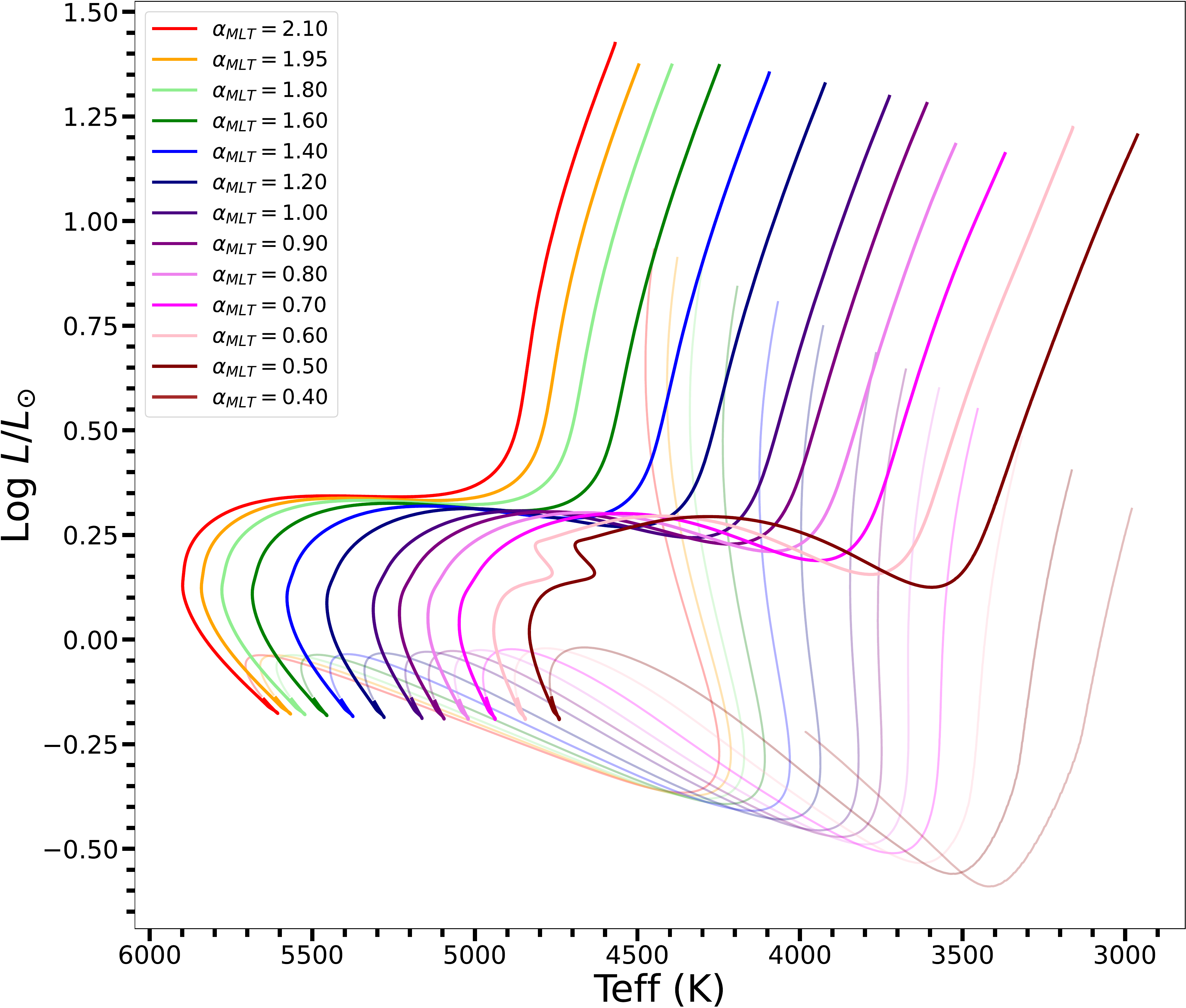}
\caption{MESA: all one solar mass, range of MLT values. Other assumptions: photosphere tables based on the MARCS model atmosphers \citep{MARCS} for atmospheric boundary conditions and the \citet{Asplund09} solar abundance scale. The pre-ZAMS evolutionary tracks are printed in fainter colors to avoid crowding and overlap with the ZAMS-RGB tracks. Note that in the case of $\alpha_\text{MLT} = 0.40$, the model failed before reaching the ZAMS.} 
\label{fig:MESA_mlt_variation}
\end{figure}  

It is well known that changes to the physical assumptions of the constituent stellar tracks will change the morphology of the isochrones, especially in mixing-sensitive regions like the main sequence turn-off (MSTO) and subgiant branch \citep[e.g.][]{Song2018, Joyce2018aNotAll, Somers2020}.
In publicly available, pre-computed isochrone databases like MIST (MESA Isochrones and Stellar Tracks; \citealt{Choi2016}), the user can specify the composition of the tracks. Occasionally, it is also possible to specify other physics, such as a degree of alpha-element enhancement (as in DSED, the Dartmouth Stellar Evolution Database; \citealt{Dotter08}), the degree of overshooting, mass loss, or inclusion of heavy element diffusion (as in BaSTI: Bag of Stellar Tracks and Isorhcones; \citealt{BaSTI2021}), and/or whether the underlying models were rotating (e.g.\ MIST). Synthetic photometry for many observational systems is also available in many isochrone databases.

However, there is no publicly available isochrone database which permits the variation of $\alpha_\text{MLT}$. Figure \ref{fig:MESA_isochrones_logL} shows the effect of variation in $\alpha_\text{MLT}$ for a set of isochrones whose ages, compositions, and other physical assumptions are otherwise identical. As Figure \ref{fig:MESA_mlt_variation} would suggest, the impact of lowering $\alpha_\text{MLT}$ in the constituent stellar tracks propagates into the isochrones, with ultra-low-$\alpha_\text{MLT}$ models exhibiting a convective core at lower masses than would be expected under typical physical assumptions. We likewise observe the shift towards both lower effective temperatures and greater temperature sensitivity with decreasing $\alpha_\text{MLT}$.

\section{Solar Calibration of $\alpha_{\text{MLT}}$}
\label{sec:solar_calibration}
Because the mixing length has neither an observable counterpart in real stars nor an analog in 3D convection simulations, we must "guess" its value.
Canonically, the value of the mixing length is determined by a method known as \textit{solar calibration} \citep[e.g.][]{CharbonnelLebreton1993}. In this process, $\alpha_\text{MLT}$ is iteratively adjusted in a solar model until the model's temperature, luminosity, radius, and other observationally constrained features (e.g $p$-mode asteroseismic spectrum) are reproduced to precisions of (ideally) at least 1 part in $10^5$ at the solar age. Historically, the Sun was the only star with a sufficient number of independent observational constraints and sufficient precision on those constraints to make this calculation feasible.    

Differences in both algorithms and physical assumptions across stellar evolution codes mean that the "solar-calibrated mixing length" is not a universal concept. The value of $\alpha_{\text{MLT},\odot}$ must be independently determined within each stellar evolution code and again for each set of physical assumptions. For example, $\alpha_{\text{MLT}, \odot}$ will, in general, be higher for solar models that incorporate heavy element diffusion than for solar models that do not, and differ at the level of at least $\sim10$\% depending on choice of atmospheric boundary conditions. 
Table 1 of \citet{Joyce2018aNotAll} and Section 1 of \citet{viani2018} provide clear demonstrations of the degree of variance in $\alpha_{\text{MLT}, \odot}$ in the context of different input physics within the same stellar evolution code (DSEP and YREC, respectively). Similarly, \citet{Cinquegrana22solarcal} demonstrate that the difference between using a ``default'' solar value (calibrated to some arbitrary set of input physics) as compared to using $\alpha_{\text{MLT}, \odot}$ from a solar calibration performed for a particular set of physical assumptions relevant to the science case (in this case, the AESOPUS opacities and predictions for AGB stars) can be significant. The \texttt{simplex\_solar\_calibration} functionality in MESA is designed to make this tedious but important calculation easier for users.  

\section{Non-solar Calibrations}
\label{sec:nonsolar_calibrations}

While the Sun remains our best source of high-precision observational constraints and always will, the Sun is not an appropriate proxy for all stars. For the same reason, it is equally inappropriate to adopt the solar-calibrated mixing length \textit{ad hoc} in any given stellar model. Until recently, however, there has simply not been an empirically motivated alternative.  

\begin{figure}[H]
\includegraphics[width=0.48\columnwidth]{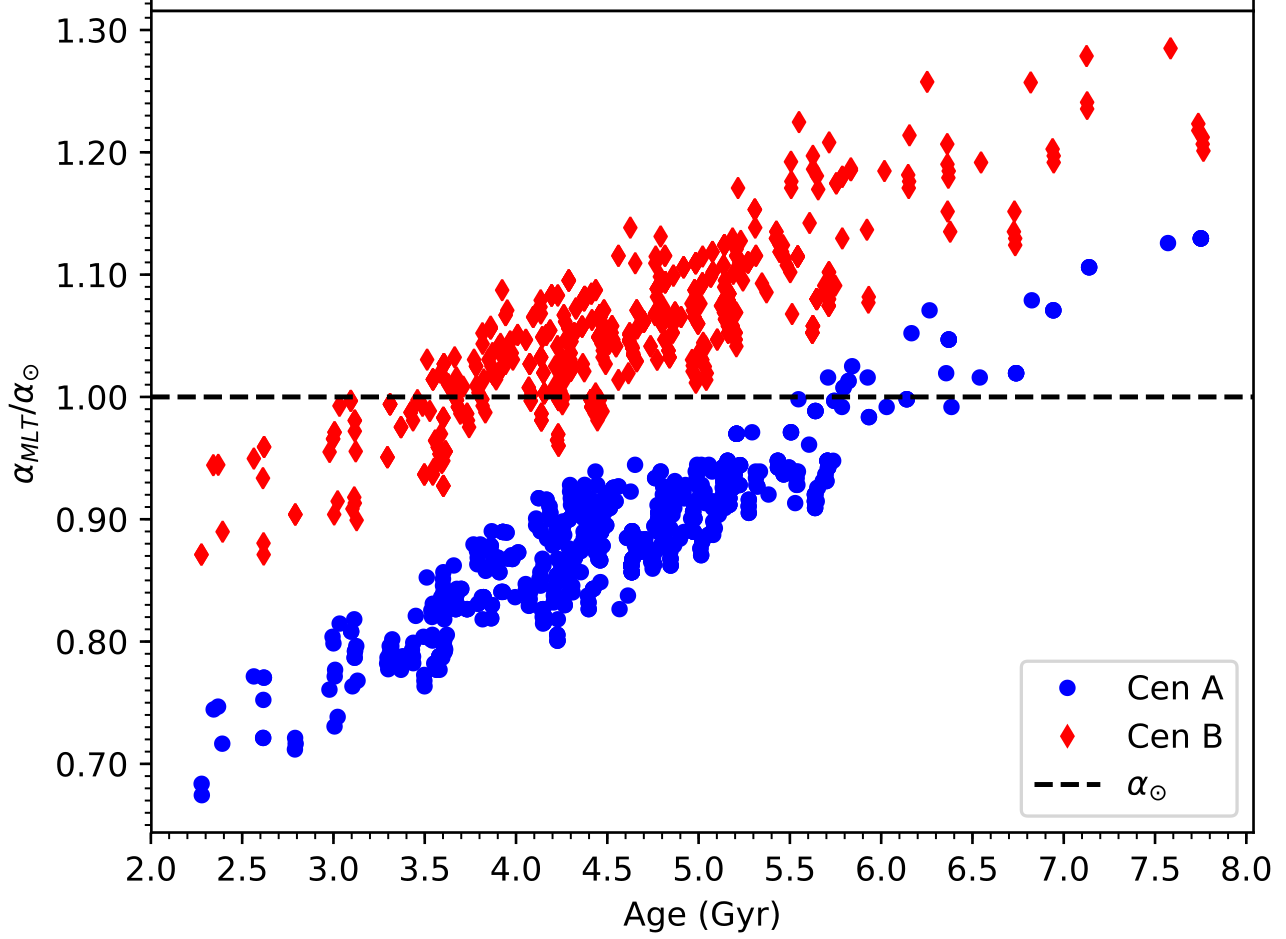}
\includegraphics[width=0.48\columnwidth]{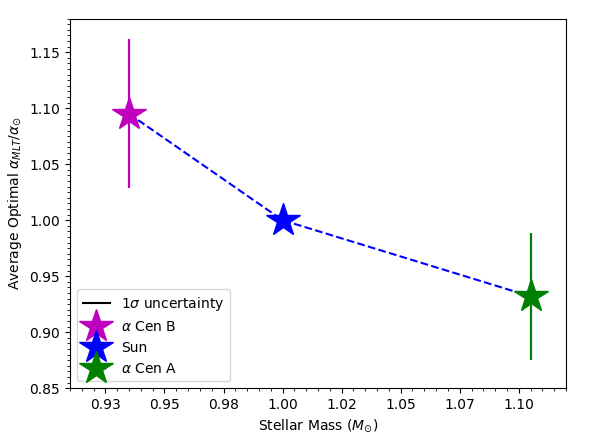}
\caption{\textbf{LEFT:} Solar-normalized $\alpha_\text{MLT}$ fits to alpha Cen A and B. This panel appears as Figure 3 in \citet{Joyce2018balphaCen} and is reproduced with permission. \textbf{RIGHT:} Optimal mixing length as a function of mass for alpha Cen A, B, and the Sun. }
\label{fig:alphaCen}
\end{figure}

Over the past few years, the identification of a handful of other stars with an appropriate number of independent observational constraints has opened the door to empirically calibrating the mixing length to non-solar targets. Due in particular to the availability of dynamical masses and robust asteroseismic information, the most reliable non-solar mixing length calibrations have been performed for alpha Centauri A and B (\citealt{GD2000, Nsamba} and \citealt{Joyce2018balphaCen}). The left panel of Figure \ref{fig:alphaCen} is a reproduction from \citet{Joyce2018balphaCen}, who found that, independent of other modeling choices, the optimal mixing length for $\alpha$ Cen B was always larger than the optimal mixing length for $\alpha$ Cen A in two-component models of the system, and in the majority of cases, the solar-calibrated value fell between them.  

Less definitive but equally compelling work has been done on non-solar mixing length calibrations to very metal-poor stars, including the work of \cite{Creevey2015} on subgiant HD 140283, whose radius is known interferometrically. The need for sub-solar $\alpha_{MLT}$ to reproduce the well-constrained temperature and luminosity of this star was later corroborated by \citet{Joyce2018aNotAll}, who also found that sub-solar $\alpha_\text{MLT}$ was necessary to fit stars of similar metal-poorness ([Fe/H] $\sim-2.4$) regardless of evolutionary phase. 

\begin{figure}[H]
\includegraphics[width=0.85\columnwidth]{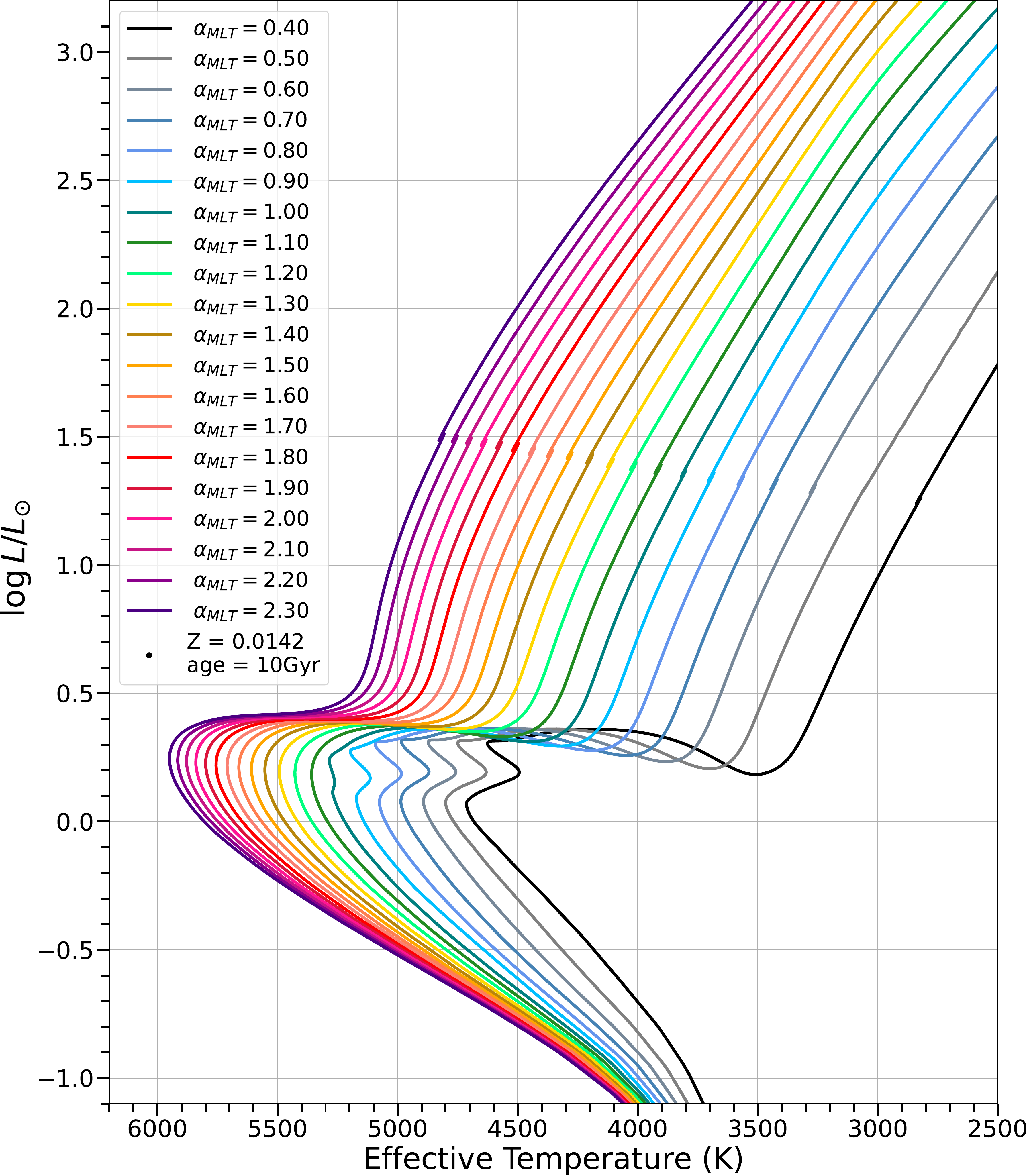}
\caption{A set of isochrones generated using MESA. All have identical compositions and ages, but the mixing length is varied. Colors of the curves correspond to the mixing length value assigned (uniformly) in the constituent tracks. }
\label{fig:MESA_isochrones_logL}
\end{figure} 

The probable dependence of $\alpha_\text{MLT}$ on metallicity has been explored by numerous observers and 1D modellers in the past decade, including \citet{Bonaca, Tayar2017, viani2018, Nsamba, Song2018} and \citet{Joyce2018aNotAll}. 
Likewise, experts in 3D radiative hydrodynamics have argued based on simulations of convection in the surface layers of stars that the mixing length should depend on composition as well as surface gravity  (e.g.\ \citealt{Freytag99, Trampedach2014, MagicMLT}).
However, the relationship between $\alpha_\text{MLT}$ and mass remains more elusive.
Notably, \citet{Kervella08} found that models of the 61 Cygni binary system ($M = 0.7$ and $0.6\,M_\odot; Z \sim Z_{\odot}$) required a sub-solar mixing length and argued that sub-solar values should be used for lower-mass stars in general. A decade later, \citet{Joyce2018balphaCen} noted a possible trend between $\alpha_\text{MLT}$ and mass, as shown in the left panel of Figure \ref{fig:alphaCen}.

\section{Scientific Applications of Changing the Mixing Length}
\label{sec:scientific_impacts}

\subsection{Implications for Age Measurements}
There are two major ways in which changes to the mixing length can dramatically impact the estimation of stellar ages, which are used for a broad range of astrophysical purposes, from identifying the evolution of planetary systems to the enrichment and merger histories of galaxies. The first is directly: changing the mixing length changes the temperature structure, which changes the nuclear burning rates, which changes the lifetime of the star. 
Generally, increasing the mixing length decreases the main sequence lifetime, while decreasing the mixing length will increase the main sequence lifetime. However, this can become more complex depending on how the model is calibrated. In a solar-calibrated model, for example, increasing the mixing length requires a {decrease in helium} to match the solar temperature at the solar age, which will then tend to {increase} main sequence lifetime back towards or even above its base length.  

The second way changing the mixing length can impact the inferred ages is less direct. As the models and their corresponding isochrones shift in temperature, the inferred age given a set of observed parameters will also shift. Concretely, one can imagine as an example a red giant star observed by Gaia, whose luminosity can be inferred from its parallax, and whose temperature and metallicity can be inferred from photometry and spectroscopy. With those parameters, one has sufficient information to fit its age using a model. However, if the model uses a solar mixing length for all stars, rather than a metallicity dependent one, we would expect the inferred mass of the star to shift by as much as 0.2 M$_{\odot}$ ($\sim$ 50 K) at [Fe/H] of $\pm$0.5, even if the solar-metallicity models are properly calibrated \citep{Tayar2017}. That shift in mass would represent a change in the inferred age of several gigayears, even as large as a factor of two in some cases. Since the shift can be metallicity dependent, it would then change the inferred age-metallicity relation of the galaxy, and fundamentally alter any inferences of its merger and enrichment history.

\begin{figure}[H]
\includegraphics[width=0.45\textwidth]{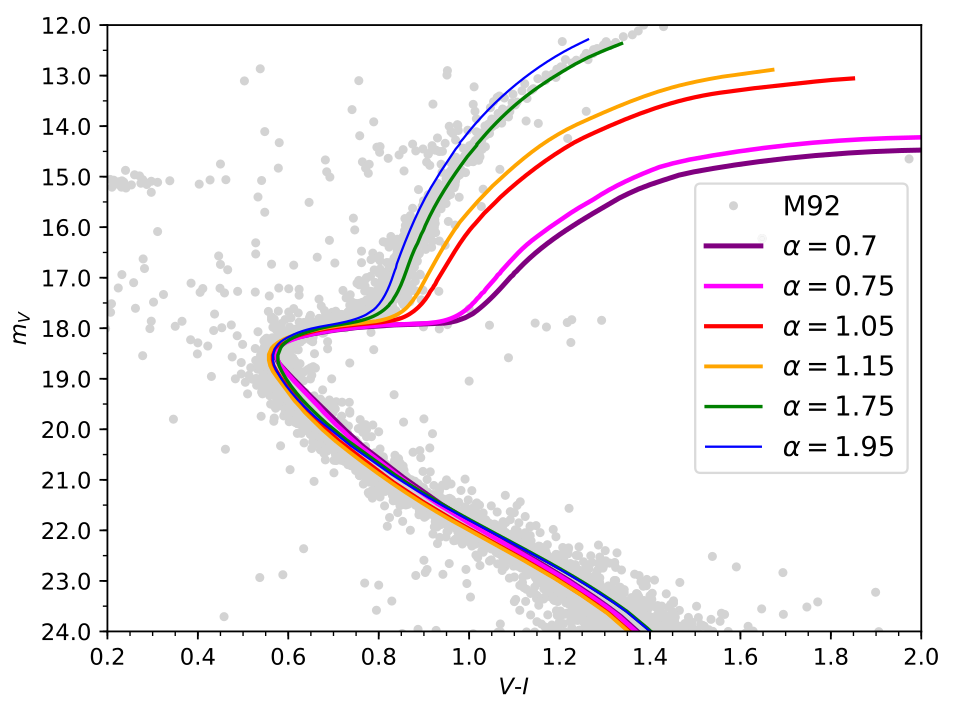}
\includegraphics[width=0.49\textwidth]{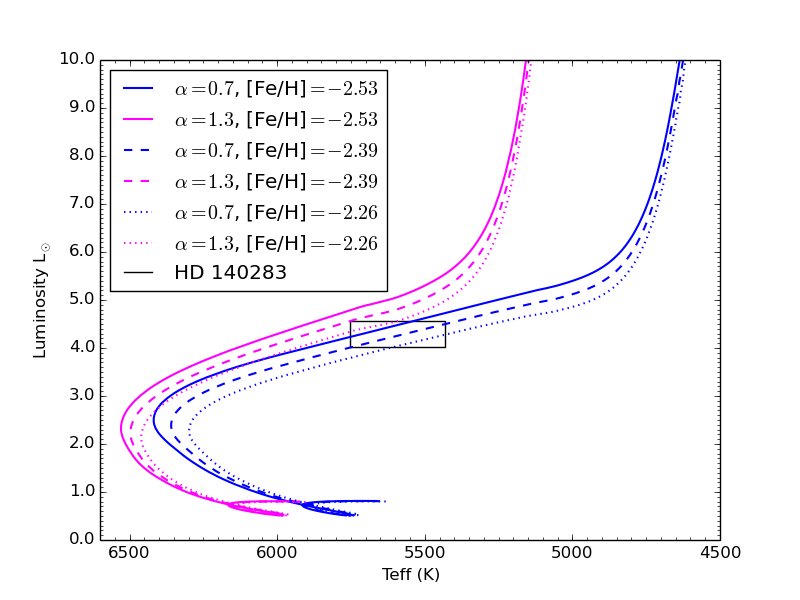}
\caption{ \textbf{LEFT:} Isochrones produced with the Dartmouth Stellar Evolution Program (DSEP) assume a range of mixing lengths. HST data for the metal-poor ([Fe/H]$\sim - 2.4$) globular cluster M92 are shown in grey. This figure appears as Figure 5 from \citet{Joyce2018aNotAll} and is reproduced with permission. \textbf{RIGHT:} Stellar tracks from DSEP designed to fit tight constraints for the metal-poor subdwarf HD 140283 are
computed with a range of assumptions about composition and $\alpha_\text{MLT}$, demonstrating that variations in one can mimic the other. This figure appears as Figure 2 in \citet{Joyce2018aNotAll} and is reproduced with permission.}
\label{fig:M92}
\end{figure}

The right panel of Figure \ref{fig:M92} demonstrates the degeneracy between metallicity and $\alpha_\text{MLT}$ using DSEP stellar tracks. The box represents brightness and temperature constraints for metal-poor subdwarf HD 140283, and the three different [Fe/H] assumptions in the tracks correspond to the minimum, maximum, and median values of the star's measured metallicity \citep{Creevey2015}. This serves as an important indication that even with excellent precision on HD 140283's metallicity, it is not possible to determine, based on this information alone, whether $\alpha_\text{MLT} = 1.3$ or $\alpha_\text{MLT} = 0.7$ provides a better fit. At the same location in luminosity--temperature space, the $\alpha_\text{MLT}=0.7$ track would provide an age estimate of order 1--2 Gyr younger than the age suggested by the $\alpha_\text{MLT} = 1.3$ curve. 

Isochrones are the models most commonly used for age determinations, especially for stellar populations. 
Figure \ref{fig:M92} shows a set of DSEP isochrones generated with various mixing lengths (colored curves) overlaid on HST photometry of metal-poor globular cluster M92 (grey points). We note in particular the sensitivity of the subgiant and red giant portions of the isochrones to the choice of $\alpha_\text{MLT}$, likewise demonstrated in the MESA-based isochrones of Figure \ref{fig:MESA_isochrones_logL}. The figure suggests that $\alpha_\text{MLT} = 1.75$ provides the best match to the data, which corresponds to a value $\sim20$\% below the solar calibration ($\alpha_{\text{MLT}, \odot, \text{DSEP}} = 1.9258$). However, we are once again presented with the problem that the effects of changing $\alpha_\text{MLT}$ are difficult to disentangle from the effects of changing the underlying composition assumptions of the isochrones---even more so than in the simpler (lower-dimensional) case of stellar tracks. As sharply demonstrated by the results of \citet{Joyce2023}, we must also reckon with the issue that the ages of isochrones are themselves degenerate with metallicity and $\alpha_\text{MLT}$, and the uncertainties on isochrone-based age determinations that do not account for this can be underestimated by a factor of two.

\begin{figure}[H]
\includegraphics[width=0.9\columnwidth]{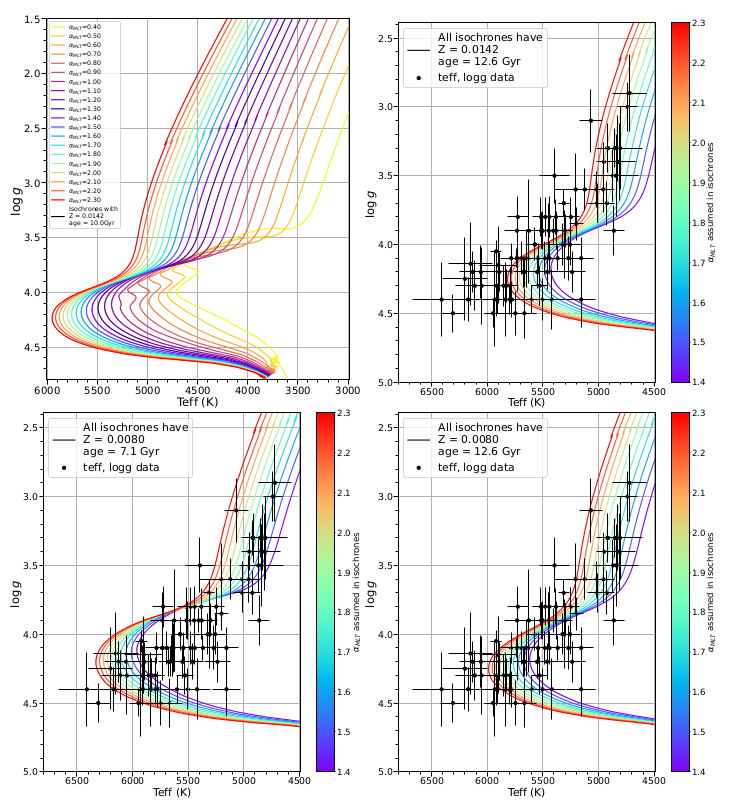}
\caption{Custom isochrones based on MESA stellar tracks assume a range of mixing length values. In the upper right and lower two panels, spectroscopic parameters for micro-lensed subdwarfs determined by \citet{Bensby17} are shown as black points with their $1\sigma$ uncertainties shown as horizontal and vertical black lines. This figure appears as Figure 11 in \citet{Joyce2023} and is reproduced with permission.}
\label{fig:isochrones_Joyce2023}
\end{figure}

The upper left panel of Figure \ref{fig:isochrones_Joyce2023} shows the same models presented in Figure \ref{fig:MESA_isochrones_logL}, but instead in the $\log g$--$T_\text{eff}$ plane. The remaining three panels show a subset of the tracks for which the mixing length assumption is varied between "reasonable extremes" for stars with masses between roughly $0.8$ and $1.0 M_{\odot}$ and metallicities from slightly super-solar to values as low as [Fe/H]$=-2.0$. Tracks in these panels sweep $\alpha_\text{MLT} = 1.4$--$2.3$, corresponding to values calibrated for similar stars in the literature (e.g.\ \citealt{Creevey2015, Tayar2017, viani2018, TangJoyce2021, Joyce2018aNotAll, Joyce2018balphaCen}). The black points show a sample of 91 micro-lensed subdwarfs with spectroscopic parameters determined by \citet{Bensby17}. The data serve to demonstrate that a physically-motivated degree of variation in $\alpha_\text{MLT}$ produces a shift in effective temperature that is at least comparable to the observational uncertainties on effective temperature, and the shift in $\log g$ is not negligible either. The upper right and two lower panels show that the spread induced by variation in $\alpha_\text{MLT}$ is present in roughly equal measure for isochrones with ages of 12.6 Gyr and 7.1 Gyr as well as for compositions near solar (Z = 0.0142) and [Fe/H]$\sim-0.3$ (Z = 0.008), demonstrating that this effect is not restricted to one particular age or metallicity regime. 

\subsection{Implications for Nucleosynthesis}

Changing the properties of convection by means of a change in the mixing length would also impact the expected nucleosynthetic processes, particularly for intermediate- and low-mass red giant and asymptotic giant branch stars. Such models also generally assume some sort of mixing length parameter, often a solar calibrated value \citep[see e.g.][]{Lugaro2012, KarakasLattanzio2014}. Increasing the mixing length parameter would increase the expected horizontal branch temperature, which then impacts the {structure, lifetime, and mixing} of AGB and super-AGB stars significantly. It has been argued that such changes can alter the predicted yields of some elements by as much as a factor of 3 \citep{Doherty2014}. It has also been found that using a different parameterization for convection, e.g.\ full spectrum turbulence rather than the mixing length, can substantially change nucleosynthetic yields \citep{Ventura20, CinquegranaKarakas2022}.

\subsection{Implications for Stars in the Instability Strip}

For some classes of pulsating stars, including $\gamma$ Doradus and $\delta$ Scuti stars, the driving of pulsations is related to the interaction between the oscillations and convection, which depends on the properties, including size, of the convective zone \citep[see][for a thorough review of this topic]{HoudekDupret2015}. Changes in the mixing length will tend to change the size of the convective envelope at a given effective temperature, and thus will change the temperatures at which stars exhibit these oscillations. Specifically,  decreasing the mixing length moves the instability strip towards lower temperature \citep{Dupret2004}. As large numbers of oscillating $\delta$ Scuti and $\gamma$ Doradus stars have been detected by recent space missions \citep{vanReeth2015, Murphy2019, Antoci_2019_puls, Aerts2023}, it has now become possible to quantitatively compare the observed instability strips to the predictions of models with different mixing lengths. Such comparisons seem to indicate that the observed instability strip is wider than that predicted by models, and one method for solving this would be to include models with a mass-dependent mixing length \citep{BowmanKurtz2018}. 

Work on the pulsation frequencies and amplitudes of classical pulsators, including Cepheids and RR Lyrae stars, have also indicated a dependence of the pulsations on the mixing length \citep[e.g.][]{Yecko1998}, and a need for a more careful treatment of the interaction between convection and pulsation than the standard mixing length theory permits \citep{Stellingwerf1982a}. The location of the red edge of the instability strip is particularly sensitive to the interactions between convection and pulsation \citep{BakerGough1967,Stellingwerf1982b,Bono1995}, and analyses have indicated that the existence of double mode pulsators may also relate to choices about convection (\citet{kollath2002} but see also \citet{SmolecMoskalik2008}). 
More generally, reducing the mixing length tends to increase the minimum mass of stars that develop blue loops and therefore pass through the Cepheid instability phase. Similarly, changing the mixing length can shift the models such that additional mass loss is required before stars can populate the RR Lyrae instability strip. As even larger catalogs of carefully studied classical pulsators are being put together at a range of metallicities \citep[e.g.][]{Jurkovic2023}, 
ongoing analysis is likely to place additional constraints on the necessary mixing lengths in these types of stars. 

\subsection{Implications for Galaxies}
Studies of unresolved stellar populations, including star clusters and other galaxies, generally estimate the stellar populations present through the comparison of observed spectrophotometry to models of stellar evolution combined to synthesize the expectation for a population \citep[stellar population synthesis, e.g.][]{Maraston1998, Maraston2005, BruzualCharlot2003, Conroy2009}. Since these models are based on stellar evolutionary tracks, they inherit the uncertainties of the underlying stellar models and their sensitivity to the choices of model physics, including the mixing length. Since changes to the mixing length cause significant shifts to the temperature of the red giant branch, and the red giant branch contributes about a third of the bolometric flux in old populations \citep{Maraston1998,Maraston2005}, changes to the mixing length can substantially change the inferred ages of galactic populations \citep{Goddard2017}. Since the mixing length may be related to stellar mass and metallicity, there are possible complexities that may be confusing our ability to infer the assembly and evolutionary history of the universe, a real issue in this time of rapidly improving data of galaxies from even earlier in time.

\section{Successes of Mixing Length}
The mixing length parameterization of convection has been astonishingly successful despite its simplicity. It has facilitated the creation of models of our own Sun as well as other stars that have propelled our  understanding of existing physics, led to the discovery of entirely new physics, and in turn given us the ability to describe the evolution of our Galaxy and Universe.

\subsection{MLT and the Standard Solar Model}
Models of stellar structure and evolution were used quite early on to comment on the possible energy generation mechanisms of stars \citep[e.g.][]{Cowling1934,Oke1950}, and even in recent history, models of the Sun have been used as a stepping stone to discover new facets of the universe. In the early 1990s, models of the Sun gave a predicted central temperature, fusion rate, and therefore neutrino flux that was inconsistent with measurements \citep[e.g.][]{Bahcall92standard}. This helped force discussion of other options for resolving the discrepancy, and eventually the neutrino oscillation model was accepted as a reasonable solution to the controversy \citep[see accounts such as][]{Perkins2014}.

More recently the absolute solar abundances have been an ongoing discussion. Spectroscopic analyses and helioseismic inversions based on solar models originally computed similar metal abundances for the Sun \citep{grevesse1993, grevessesauval1998} but the incorporation of three dimensional effects and the impact of Non Local Thermodynamic Equilibrium effects into spectroscopic analyses seemed to argue for lower overall abundances \citep{Asplund09} that are incompatible with solar modeling \citep{BasuAntia1997}. New measurements of the solar rate of production of neutrinos from the CNO cycle seem to argue for a result more consistent with the results of solar modeling \citep{BorexinoCNO}, as do some more recent estimates from spectroscopic analyses \citep{Magg2022}, and so this may again represent an instance where stellar models based on mixing length theory encouraged the shift to a new standard paradigm, with implications for a wide range of analyses. 

\subsection{MLT and Asteroseismology}

Models of stellar structure and evolution based on mixing length theory have also been sufficiently accurate to permit detailed asteroseismic analyses, allowing the study of the detailed interior profiles of other stars based on the exact frequencies and frequency ratios of their oscillation modes. While whole books can and have been written on the possibilities \citep{aerts2003} and methods \citep{BasuChaplin2018} of such analyses, we highlight here a few of the results that have enhanced the understanding of the physics of stellar interiors.  

The comparison of observations to stellar models created using mixing length theory have allowed for the identification of modes and mode patterns sensitive to the stellar interior, including to the size of the convective core, and thus to mixing in  overshoot regions \citep{vanReeth2015, Pedersen2022, Deheuvels2016, Constantino2015}. They have allowed the identification of modes of mixed character that probe the rotation profile of the interior of the star \citep{beck2011, Deheuvels2012, Mosser2012b}, and potentially identified stars that have undergone interactions \citep{Rui2021, Deheuvels2022, Li2022, Matteuzzi2023, Tayar2022c}. Models have identified modes that are missing or altered \citep{Garcia2014b,Stello2016a} and argued that these can probe the internal magnetic field strength and structure \citep{Fuller2015c, Bugnet2021}. More generally, asteroseimic analyses combined with stellar models have allowed the estimation of ages for large populations of stars \citep[e.g.][]{Pinsonneault2018} and these results have then been used to infer the evolutionary histories of our own \citep{SilvaAguirre2018} and other \citep{Chaplin2020, Grunblatt2021} galaxies, as well as served as training sets for much larger explorations \citep{Ness2016, Leung2023}.

\section{Observational Challenges of Mixing Length}
Given the remarkable success of one dimensional evolutionary models of stars and their utility for studies of everything from extrasolar planets to the populations of the earliest galaxies, one might question whether there is still a need to devote effort to discussions of the choice of $\alpha_\text{MLT}$.
{We argue that such conversations must continue. 
Large numbers of stars are now observationally characterized to levels of precision that show our simple assumptions about the mixing length generate demonstrably incorrect models.}
This has been identified through 
the careful calibration of individual stars \citep{Joyce2018balphaCen}, 
the use of populations of clusters \citep{Brasseur2010, Cohen2015, Smiljanic2016, Joyce18not}, 
constraints on other populations of stars \citep{Ness2013}, 
asteroseismic analysis of dwarf and giant stars \citep{Bonaca2012, Metcalfe2014, Creevey2016, SilvaAguirre2017, Tayar2017, viani2018}, efforts in the M dwarf regime \citep{FeidenChaboyer2012b, SomersPinsonneault2014} and work on the location of the instability strip \citep{BowmanKurtz2018}, to name a few. 

The most interesting and challenging part of the MLT landscape is that these analyses do not all agree with each other. Although each argues for a change in the mixing length to better match observations, {these changes do not converge to a single different value or scaling relation}. Perhaps there exists a relationship between the mixing length and variables like mass, metallicity, surface gravity, temperature, magnetic field, etc., that could match all of the observational constraints over all regimes, but no such single solution has yet been shown to work. It could also be the case that these discrepancies are showing us that the current mixing length framework is finally insufficient for the modern observational data climate, and there is a real need for a new paradigm. 

\begin{figure}
\begin{center}
\subfigure{\includegraphics[width=0.49\textwidth, clip=true, trim=0cm 0cm 0cm 0.8cm]{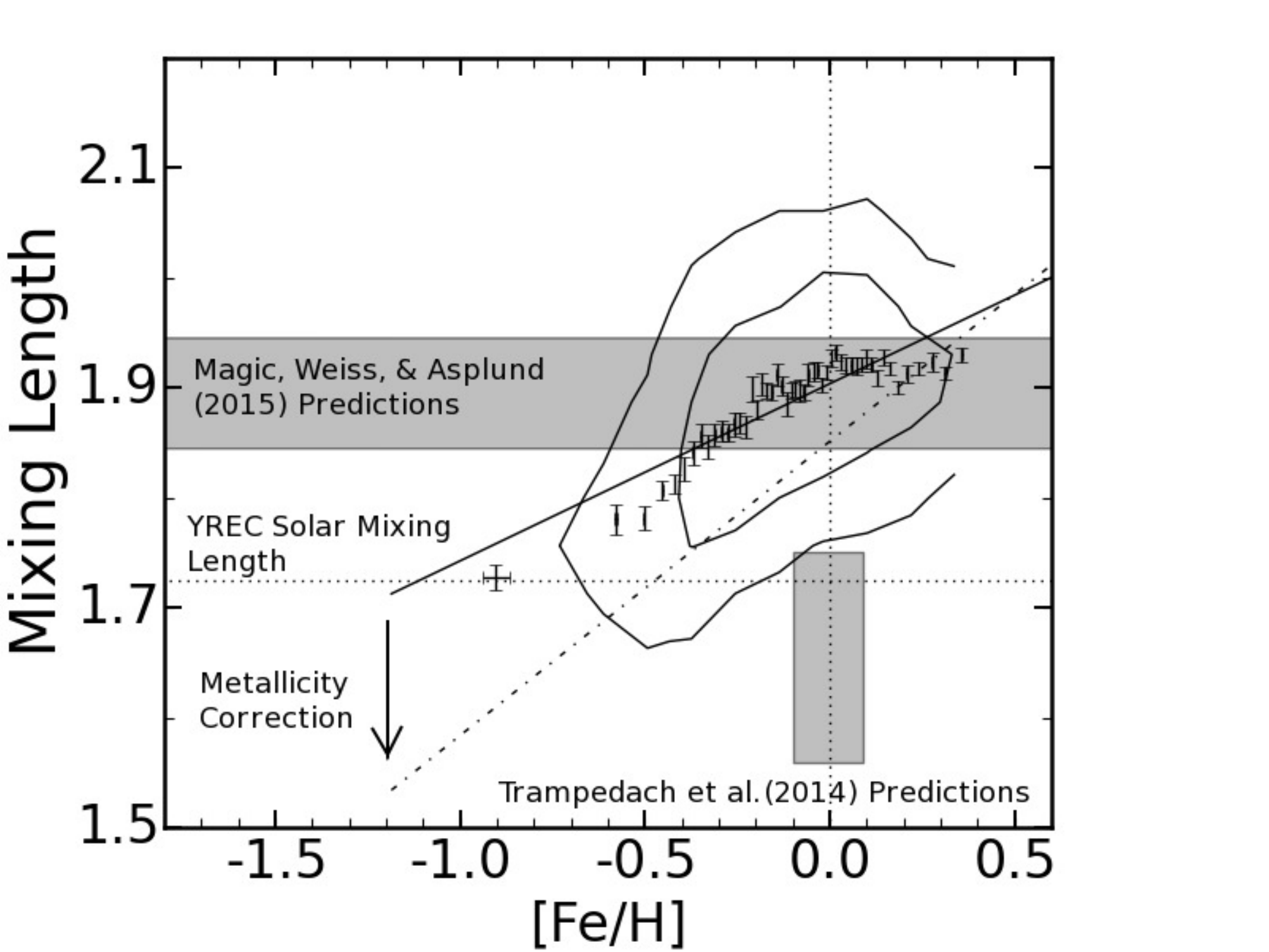}}
\subfigure{\includegraphics[width=0.49\textwidth, clip=true, trim=0cm 0cm 0cm 0.5cm]{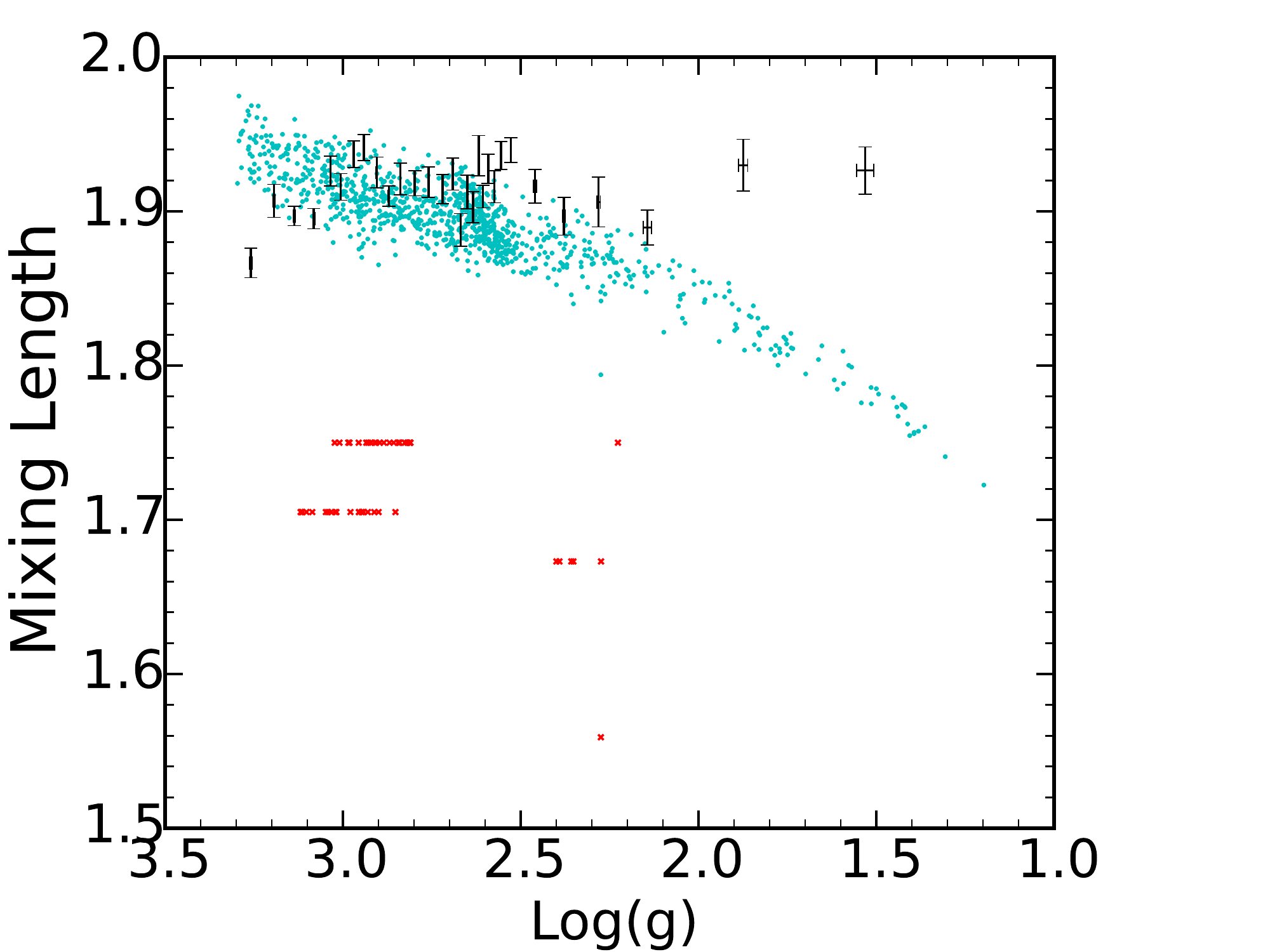}}
\caption{{\bf Left}: The effective mixing length required to match observations in a set of 1D stellar evolution models (black points) does not agree with the predictions of 3D simulations (grey bands) as a function of metallicity. {\bf Right:} The effective mixing length required to match observations in a set of 1D stellar evolution models (black points) does not agree with the predictions of 3D simulations (teal and red points) as a function of surface gravity. Reproduced with permission from \citet{Tayar2017}. }  
\label{Fig:mlt17}
\end{center}
\end{figure}

Another open question concerns the inability for more sophisticated three dimensional simulations to predict a mixing length that matches the requirements from the observations \citep[See e.g.\ Figure \ref{Fig:mlt17}, reproduced from][]{Tayar2017}. The translation from a three dimensional simulation to some estimate of a mixing length is not straightforward \citep{Trampedach2014,MagicMLT}, since there is no clear physical definition for such a parameter, so the lack of correspondence could be a translation issue rather than evidence of missing physical understanding. Three-dimensional simulations of convection also have their own challenges \citep{Miesch2005}, as do observational calibrations to place stars on a fundamental scale \citep{Tayar2022}. Even so, the inability to predict the mixing length from well known physical principles calls into question its long-term viability as a relevant model.

\section{The Future of MLT} 
Having reviewed both the extraordinary successes
and substantial limitations of the mixing length theory of convection, we now speculate on what the future holds for MLT. 

First and foremost, we expect the evolution of MLT to be driven observationally. As both data quality and data volume increase at unprecedented rates, the data will require models of a higher standard to make sense of abundant, high-precision observational constraints from multiple sources and to extend the interpretation of data to regimes where the data is not yet sufficient to constrain the model.

{On the five- to ten-year horizon, we can expect to see empirical calibrations of $\alpha_\text{MLT}$ to an increasingly broad, globally representative sample of non-solar targets, and for these data-driven values to increasingly replace \textit{ad hoc} use of the solar prescription}.
Carefully characterized targets with well-constrained masses, compositions, and radii will be assembled into a list of calibration targets used to determine the appropriate mixing length for an individual model grid; we expect these targets to come from a combination of eclipsing binary analysis, open cluster membership, and/or photometric, astrometric, spectroscopic, and asteroseismic data.  
{Grids of models that span various physical assumptions will be generated to estimate the appropriate mixing length for each star's particular composition, temperature, and surface gravity. Arguments will be made about whether a single mixing length should then be used to fit \textit{all} stars with mass $X$ and/or metallicity $Y$, whether there is a scaling relation between $\alpha_\text{MLT}$ and other stellar parameters that should be applied universally or in particular evolutionary phases/mass regimes, or  whether models should be using a static mixing length, rather than one that changes over time, at all. Comparisons will be made to three-dimensional models to determine if they can be processed in such a way as to provide useful insight on the changes to the mixing length.} Stellar evolution codes will be modified to accommodate such schemes, and, over time, we will learn which approaches are better.

We expect that this paradigm {of iterative, data-informed revision and improvement} will persist for a time, and then the data will once again improve. At that point, it will become more clear whether the calibrated mixing length framework is sufficient to predict the evolution, temperature, and ages of stars to an acceptable precision. 
When that happens, the researchers of the future may finally have to find an implementation of convection with greater physical fidelity
to use in stellar evolution calculations. 
Or, they may once again discover that the humble mixing length prescription once again requires only slight modifications to do a truly excellent job predicting the behavior and evolution of stars across the Hertzsprung--Russell diagram. 
In either case, researchers will be driven to make choices that preserve stellar models' place as a fundamental pillar of astrophysics.

\section*{Acknowledgements}
The authors wish to thank L\'aszl\'o Moln\'ar for sharing his expertise in classical stellar variability and Giulia Cinquegrana for sharing her expertise in TP-AGB stars, both of which contributed to this manuscript. The authors also wish to thank John Bourke for revisions and discussion. 

M.J. gratefully acknowledges funding of MATISSE: \textit{Measuring Ages Through Isochrones, Seismology, and Stellar Evolution}, awarded through the European Commission's Widening Fellowship.  
This project has received funding from the European Union's Horizon 2020 research and innovation programme.

%%%%%%%%%%%%%%%%%%%%%%%%%%%%%%%%%%%%%%%%%%
%\bibliographystyle{aasjournal}
\bibliography{mlt_review,mlt_review2,library,library2,tayar2017, Giulias_bib}
\end{document}